\documentclass[prb,superscriptaddress]{revtex4}

\usepackage{amsmath}                   
\usepackage{amsfonts}                  
\usepackage{graphicx}                 

\begin{document}

\author{Alexey A. Polotsky}
\email{alexey.polotsky@gmail.com}
\affiliation{Institute of Macromolecular Compounds, Russian Academy of Sciences 31 Bolshoy pr., 199004 St.-Petersburg, Russia}

\author{Mohamed Daoud}
\affiliation{Service de Physique de l'Etat Condens\'e CEA Saclay, 91191 Gif-sur-Yvette Cedex, France}

\author{Oleg Borisov}
\email{oleg.borisov@univ-pau.fr}
\affiliation{Institute of Macromolecular Compounds, Russian Academy of Sciences 31 Bolshoy pr., 199004 St.-Petersburg, Russia}
\affiliation{Institut Pluridisciplinaire de Recherche sur l'Environnement et les Mat\'eriaux, UMR 5254 CNRS/UPPA, Pau, France}

\author{Tatiana M. Birshtein}
\affiliation{Institute of Macromolecular Compounds, Russian Academy of Sciences 31 Bolshoy pr., 199004 St.-Petersburg, Russia}

\title{A Quantitative Theory of Mechanical Unfolding of a Homopolymer Globule} 

\begin{abstract}
We propose the quantitative mean-field theory of mechanical unfolding of a globule formed by long flexible homopolymer 
chain collapsed in poor solvent and subjected to extensional deformation. We demonstrate that depending on the degree of 
polymerization and solvent quality (quantified by the Flory-Huggins $\chi$ parameter) the mechanical unfolding of the collapsed 
chain may either occur
continuously (by passing a sequence of uniformly elongated
configurations) or involves intra-molecular micro-phase coexistence of a collapsed and a stretched segment followed by an abrupt unravelling transition. 
The force-extension curves 
are obtained and quantitatively compared to
our recent results of numerical self-consistent field (SCF) simulations. The phase diagrams for extended homopolymer chains
in poor solvent comprising one- and two-phase regions are calculated
for different chain length or/and solvent quality.   
\end{abstract}
\maketitle
%
\section{Introduction} \label{Sec:Intro}
Recent developments in polymer micromanipulation techniques such as atomic force microscopy (AFM) or optical tweezers have made it possible 
to subject an individual polymer molecule to mechanical deformation and to investigate the response of the molecule to this 
stimulus~~\cite{Rief:1997, Hugel:2001, Haupt:2002, Smith:1996, Kellermayer:1997}. 
The output provided by such single molecule force spectroscopy experiments is commonly expressed in the form of the force-extension curves.  
The ultimate goal of theory of micromechanical deformation of macromolecules is to rationalize the shape of the force-extension curves obtained 
in experiments, i.e. to correlate different patterns (regimes) on the force-extension curves with specific conformational 
transitions that occur upon mechanical action on the macromolecule.

The most complex patterns (non-monotonic, oscillating dependences) in the force-extension curves are observed upon mechanical manipulations with biological polymers, i.e., globular proteins, DNA and
polysaccharides;  these patterns are attributed to cascades of intra-molecular conformational transitions related to unfolding of 
hierarchically organized intra-molecular structures which are stabilized by strong intra-molecular attractive forces operating in aqueous media. However, even such a simple system as a homopolymer globule collapsed in poor solvent may exhibit a non-trivial behavior under extensional deformation. Analysis of this model system is highly important for rationalization of the relationships of mechanical unfolding of more complex globular structures and supra-molecular assemblies of
biomacromolecules.

There exist two modes of micromechanical manipulation depending on whether the value of deformation (extension) $D$ or that of the force $f$ applied to the macromolecule is chosen as the control parameter~\cite{Skvortsov:2009}. From the point of view of the statistical mechanics, one can consider these two situations as corresponding to different statistical \emph{ensembles}: 
The former case will be referred to as the \emph{constant extension ensemble}, or $D$-ensemble, whereas the latter - as the \emph{constant force ensemble}, or $f$-ensemble.

\begin{figure}[t] 
  \begin{center}
  \includegraphics[width=12cm]{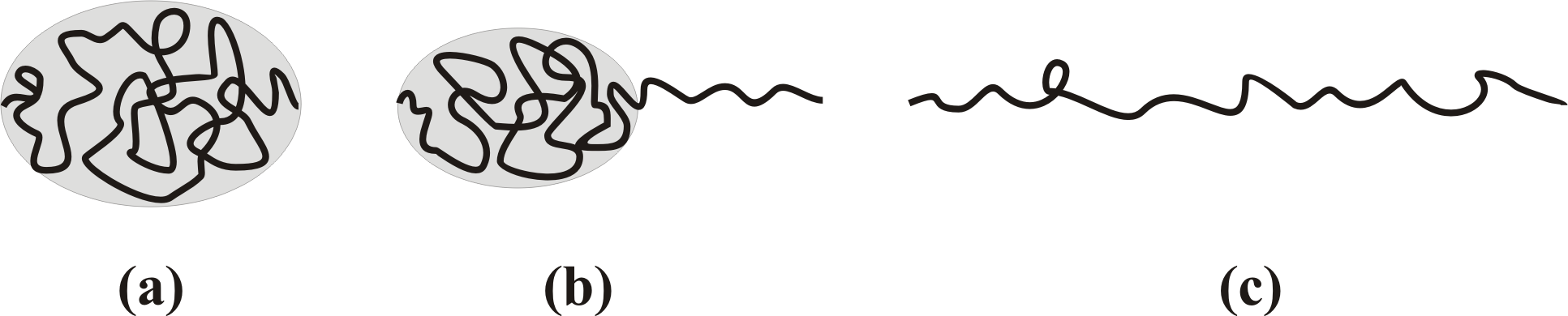}
  \end{center}
  \caption{Conformation of stretched globule: prolate (ellipsoidal) (a), tadpole (b) and stretched chain (c).}
  \label{fig:globules}
\end{figure}

The first theory of globule unfolding in the constant extension ensemble was developed 
by Halperin and Zhulina~\cite{Halperin:1991:EL} in the framework of the scaling approach.
This theory is strictly applicable close to the theta-point since the solvent strength is accounted for in scaling terms 
and only for long chains comprising large number of thermal blobs. 
It was demonstrated that there exist three regimes of globule deformation: 
(i) At weak extensional deformations the globule has an elongated shape, Figure \ref{fig:globules}~b, 
the reaction force grows linearly with extension. 
(ii) The further increase in the end-to-end distance leads to appearance of a new intra-molecular micro-phase, i.e., 
a (strongly) stretched chain part. 
In a wide range of 
extensions, a microphase segregation takes place and the globule acquires a ``tadpole'' conformation with a prolate globular ``head'' 
and a stretched ``tail'', Figure \ref{fig:globules}~b. 
The deformation  is accompanied by progressive unfolding of the globular core and occurs at almost constant reaction force. 
This be\-ha\-vior has an analogy with the Rayleigh instability in a liquid droplet~\cite{Rayleigh:1882} under additional constraint of 
connectivity of solvophobic monomer units in the chain. (iii) At strong deformations the tadpole conformation becomes unstable 
and the globule completely unfolds to a uniformly extended chain,
Figure \ref{fig:globules}~c. The reaction force grows again in this regime.
According to the theory of Halperin and Zhulina, 
transitions between the regimes occur continuously upon an increase in the end-to-end distance $D$; in particular, 
the tadpole regime ends up when the size of the tadpole head becomes comparable to the thermal blob size (equal the thickness of the tadpole tail) or, in other words, when the head becomes indistinguishable from the tail. 

Later, Cooke and Williams~\cite{Cooke:2003} considered stretching of a dry (solvent free) globule. They found that the 
tadpole conformation becomes unstable when the head of the tadpole still contains a large number of monomers, $n \sim N^{3/4}$ 
(in contrast to $n \sim 1$ that follows from prediction of the theory of Halperin and Zhulina~\cite{Halperin:1991:EL}, 
if it is extrapolated to the dry globule case). This ``unraveling transition''~\cite{Cooke:2003} (i.e. complete globule unfolding) is accompanied by a sharp drop in the reaction force. 


Craig and Terentjev~\cite{Craig:2005:1} extended the Cooke and Williams approach to the case of 
collapsed semiflexible polymers and took into account chain stiffness and finite extensibility. 
Depending on the persistence length the polymer collapses into spherical (small persistence length) or toroidal 
(large persistence length) globule. It was shown that unfolding of both spherical and toroidal globule occurs with a decay in the 
reaction force in the tadpole regime (the tadpole's head is correspondingly spherical or toroidal) with the following drop when the 
globule completely unravels. In addition, Craig and Terentjev also considered the model of ordered globule which can describe unfolding of inhomogeneous globules.

In our recent work~\cite{Polotsky:2009}, we have performed a self-consistent field (SCF) modeling of equilibrium unfolding of a homopolymer globule
subjected to extensional deformation, using Scheutjens-Fleer numerical SCF (SF-SCF) approach~\cite{Fleer:1993}. 
This method has allowed us to obtain conformational and thermodynamic properties of the considered system (e.g. the force-extension curves) 
in a wide range of polymerization 
degrees and solvent qualities and to reveal a general picture of the globule deformation in the constant extension ensemble. 
It was found that the globule unfolding occurs in three stages (Figure \ref{fig:globules}), 
in accordance with the prediction of Halperin and Zhulina~\cite{Halperin:1991:EL}. 
Our simulations indicate that the transition between weakly stretched and tadpole states is continuos 
whereas the transition from the tadpole to the uniformly  stretched chain occurs junpwise and is accompanied by the force drop,
in accordane with prediction of Cooke and Williams~\cite{Cooke:2003}. 
On the other hand, we have shown that for short chains, under moderately poor solvent strength conditions, 
unfolding occurs continuously: 
upon an increase in the imposed end-to-end distance the extended globule retains a longitudinally uniform shape at any degree of deformation.

The aim of the present paper is to develop a quantitative analytical theory of globule unfolding in the constant extension ensemble 
that explicitly takes into account the (poor) solvent quality and enables us to analyze in details effects of finite chain length.
This theory will allow us to go beyond the capacity of the SCF modeling and to calculate  
the globule-tadpole and the tadpole-stretched chain transition points (i.e. to construct the phase diagram) 
and to find corresponding conformational characteristics, 
reaction forces and force jump at transition, in a wide range of polymerization degree $N$ and  solvent quality expressed via 
Flory-Huggins parameter $\chi$. 
In contrast to earlier theories our model explicitely accounts for the induced asymmetry of the globule in the phase coexistence regime 
(we remark that the prolate shape of the tadpole head was recently found by Grassberger and Hsu~\cite{Grassberger:2002} 
in their analysis of the Monte Carlo simulation data). 
Using the developed theory we perform a comprehensive analysis of the globule unfolding that includes 
(i) calculation of force-extension curves and their quantitative comparison with those obtained in SCF modeling~\cite{Polotsky:2009}, 
(ii) a detailed study of the phase transition, construction of the system's phase diagram, calculation of globule characteristics in 
the transition points (iii) derivation of asymptotic analytical dependences for characteristics of deformed globule in different 
deformation regimes and in the transition points.

It should be emphasized that both the theory we develop as well as the SCF approach used in~\cite{Polotsky:2009} are of the \emph{mean field} type. It is well known that the mean-field approach neglects fluctuations around the ground state for the considered system~\cite{GrosbergKhokhlov:1994}. This approximation is justified for large systems except of the vicinity of the phase transition points, but may fail for ``small'' systems.

Although the kinetic aspects of the globule unfolding lie beyond the scope of the present paper, 
let us note that in real unfolding experiments and in computer experiments the distance between chain ends is increased 
continuously at some constant finite velocity. The extension rate plays an important role and may determine the choice of the unfolding pathway (see, for example~\cite{Cieplak:2004}). From this point of view, our analysis corresponds to an infinitely slow globule deformation such that at each end-to-end distance $D$ the system has enough time to reach the equilibrium state. 
However, our theory may be insightful for the experiments in which the unfolding kinetics is studied because it allows 
to estimate the height of the barrier separating different states (different minima) on the free energy landscape.

The rest of the paper is organized as follows. 
In the section ``Model and general formalism'' we introduce our model and derive free energies 
for weakly elongated globule, for strongly extended (unfolded) chain
and for the microphase segregated (tadpole) state. 
The results of the calculations, including force-extension curves, system's parameters in the transition points and phase diagrams are summarized 
in the sections ``Numerical Results'' and ``Analytical Results'' that is followed by Discussion and Conclusions.

%

\section{Model and general formalism} \label{sec:model}
\subsection{Unperturbed and weakly extended globule} \label{subsec:ellipsoid}

\subsubsection{Unperturbed globule}
Let us consider a flexible polymer chain comprising $N$ monomer units, each of size $a$, immersed in a poor solvent.
The solvent strength is charaterized by the Flory-Huggins parameter $\chi \geq 0.5$. 

Following Ref.~\cite{GrosbergKhokhlov:1994} we consider the polymer globule in the "volume approximation", that is the globule is envisioned as a ``liquid droplet'' 
of a constant number density of the monomer units $\varphi$
\begin{equation}
	\varphi = \frac{N}{V} ,
\end{equation}
where $N$ is the number of monomers in the globule and $V$ is its volume. 
The correlation length of the density fluctuations in the globule $\xi\sim \varphi^{-1}a^{-2}$ is assumed to be much smaller that the size of the 
globule $R_{globule}\sim V^{1/3}$ that is the case either for large $N$ or for sufficiently poor solvent strength conditions.
The globule free energy can be represented in the form
\begin{equation} \label{eq:Fglobule}
	F_{globule} = \mu N + \gamma S ,
\end{equation}
where $\mu$ is the monomer chemical potential (free energy) in \emph{infinite} globule ($\mu$ is negative) and $\gamma$ 
is the interfacial tension coefficient ($\gamma$ is positive). 
Here and below $k_BT$ is taken as energetic unit.

Moreover, as has been demonstrated~\cite{Lifshitz:1978, GrosbergKhokhlov:1994},
the conformational entropy of the chain in the globular state also
scales proportionally to the globule surface area. Hence, 
$\gamma$ also comprises a contribution of the conformational entropy to the globule's free energy. 

Due to the interfacial tension, a free (unperturbed) globule has the shape of a sphere with the radius
\begin{equation} \label{eq:R_0}
	R_0 = \left(\frac{3N}{4\pi\varphi}\right)^{1/3}
\end{equation}
and interfacial area $S_0 = 4\pi R_0^2$.

Within volume approximation all the partial parameters~\cite{Flory:1953} $\varphi$, $\mu$, and $\gamma$ which characterize the globular state 
do not depend on the number of monomers units in the globule $N$ and are solely determined by the solvent quality 
(i.e. by the Flory-Huggins parameter $\chi$), that is 
$$
\varphi   =  \varphi(\chi) 
$$
$$
\mu  =  \mu(\chi) 
$$
$$
\gamma  =  \gamma(\chi)
$$


The dependence of $\varphi$ and  $\mu$ on $\chi$ can be easily found in the framework of the Flory lattice model 
of polymer solution~\cite{Flory:1953, deGennes:1979}. For a solution of chains with the polymerization degree $N$ and the volume fraction $v\equiv\varphi a^3$, the free energy \emph{per lattice site} is given by
\begin{equation}  \label{eq:Fpersite}
    F_{site}=\frac{v}{N}\log v + (1-v)\log(1-v) + \chi v (1-v).
\end{equation}
If we cancel the first term in Eq.~(\ref{eq:Fpersite}) corresponding to translational entropy of polymer chains in solution, 
we obtain the expression for the free energy of the solution comprising one infinitely long chain.

The equilibrium volume fraction $v$ of the monomer units in the globule (which is found in equilibrium with pure solvent) 
is determined from the condition of vanishing 
osmotic pressure $\pi$
\begin{equation}  \label{eq:Osmotic}
    \pi = v^2 \, \frac{\partial }{\partial v} \left( \frac{F_{site}}{v} \right)
    = 0.
\end{equation}
This leads to
\begin{equation}  \label{eq:chi_tau}
  \chi = -\frac{\log(1-v)}{v^2}  - \frac{1}{v},
\end{equation}
i.e. the desired dependence $\varphi=\varphi(\chi)\equiv a^{-3} v(\chi)$ is obtained in ``inverse'' form (we remark that 
the translartional entropy is eliminated from the latter equation).

On the other hand, $\mu$ amounts to the monomer free energy change when it is transferred from the pure solvent (dilute phase, which is taken as a reference state for calculation of the chemical potential) 
to the globular phase with the concentration $\varphi$. 
Therefore, first we re-write the free energy (\ref{eq:Fpersite}) per monomer:
\begin{equation}  \label{eq:Fpermon}
    F_{monomer} = \frac{F_{site}}{v} = \frac{1-v}{v}\log(1-v) 
                 + \chi (1-v),
\end{equation}
then
\begin{equation}  
    \mu \equiv F_{monomer}(v) - F_{monomer}(v \to 0) =
    -\chi v + 1 + \frac{1-v}{v}\log(1-v)
\end{equation}
or, using Eq. (\ref{eq:chi_tau})
\begin{equation}  \label{eq:gamma1_tau}
  \mu = 2 + \frac{2-v}{v}\log(1-v).
\end{equation}

Finding the dependence of $\gamma$ on $\chi$ is a more difficult problem that has not a complete analytical solution yet. 
However, for moderately poor solvent (close to the coil-globule transition point) where the polymer-solvent interaction free energy can be represented in terms of virial expansion $F_{int, site}=\frac{B}{2}\cdot v^2 + \frac{C}{6}\cdot v^3$, $B$ and $C$ are second and third virial coefficients, respectively, a closed form analytical solution for $\gamma$ can be obtained in the framework of the Lifshitz theory of polymer globules~\cite{Lifshitz:1978, GrosbergKhokhlov:1994}: 
\begin{equation} \label{eq:gamma:virial}
  \gamma a^2 = \frac{3B^2}{16 C^{3/2}}
\end{equation}
Exact numerical prefactor in Eq.~(\ref{eq:gamma:virial}) was calculated by Ushakova \emph{et.~al.}~\cite{Ushakova:2006} (see Eq. (16) and the following discussion in Ref.~\cite{Ushakova:2006}). In our case, expansion of the free energy (\ref{eq:Fpersite}) at small $v$ gives $B = (1 - 2\chi)$ , 
$C = 1$ and, therefore.
\begin{equation} \label{eq:gamma_chi}
  \gamma a^2 = \frac{3}{16} (1-2\chi)^2.
\end{equation}
 
On the other hand, the values of $\gamma$ as well as those of $\varphi$  and $\mu$ as functions of $\chi$ can be found using the numerical SCF approach which was successfully applied to model the globule unfolding in our previous work~\cite{Polotsky:2009}. The method of finding the parameters $\varphi$, $\mu$, and $\gamma$ are described in Appendix, the numerical values of $\varphi$, $\mu$, and $\gamma$ for $\chi$ values ranging from $\chi=0.8$ to $2$ are given in Table 1 of Appendix. These values are also plotted in Figure \ref{fig:mugammaphi} together with $\varphi(\chi)$, $\mu(\chi)$ and $\gamma(\chi)$ given by Eqs~(\ref{eq:chi_tau}), (\ref{eq:gamma1_tau}) and (\ref{eq:gamma_chi}), respectively. We see a very good agreement of the numerical results for $\varphi$ and $\mu$ with the Flory theory in the whole range of $\chi$, whereas for $\gamma$ an agreement is observed only at small $\chi$ (i.e. where the virial expansion is applicable). Moreover, for strong solvent ($\chi > 1.2$) we see that $\gamma$ has a scaling on $(1 - 2\chi)$ different than the quadratic one predicted by Eq. (\ref{eq:gamma_chi}). Therefore, in the following numerical calculations we will use the values of  $\varphi$, $\mu$ and $\gamma$  given in Table 1.

\begin{figure}[t] 
  \begin{center}
  \includegraphics[width=10cm]{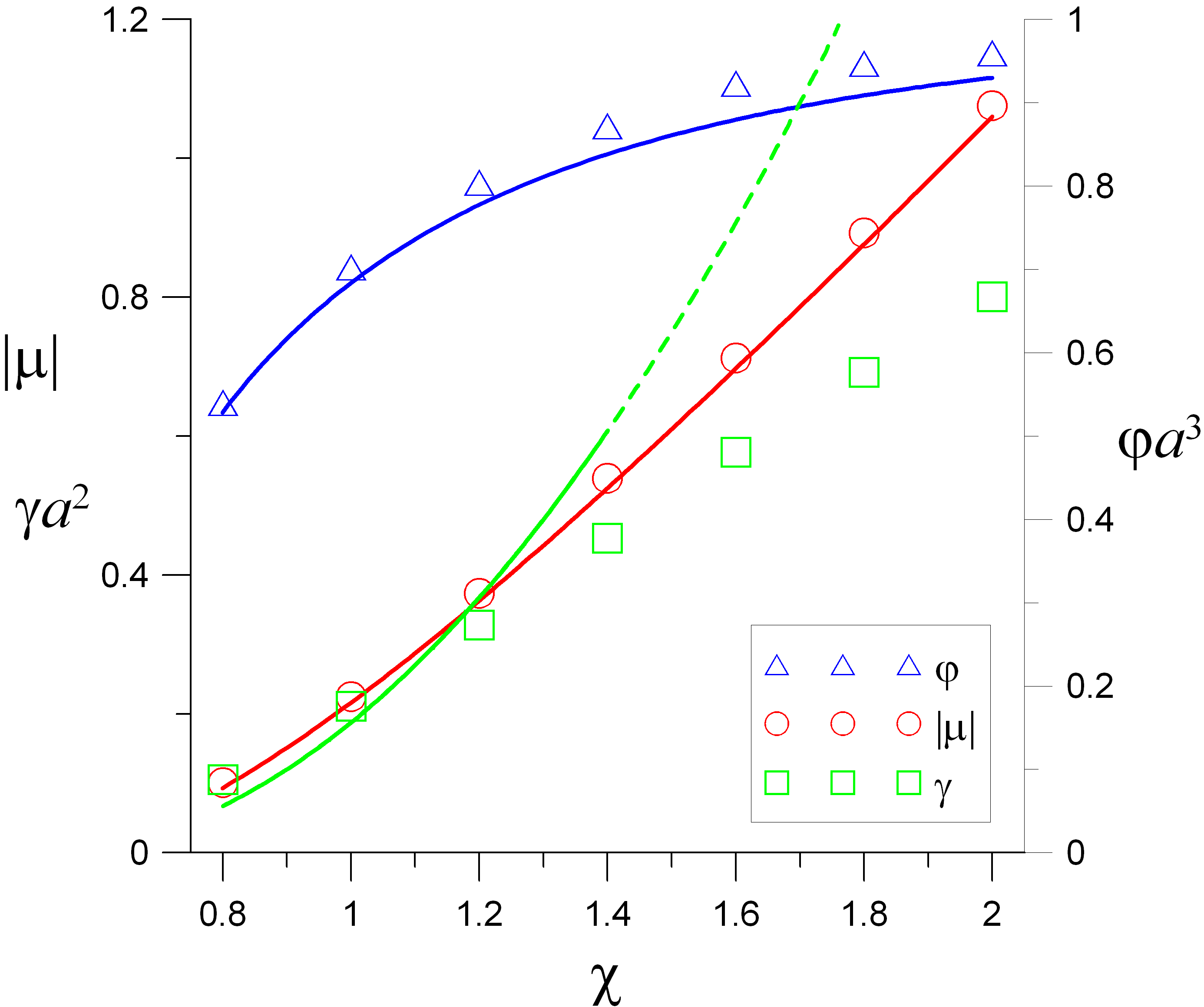}
  \end{center}
  \caption{Monomer chemical potential $\mu$, interfacial tension coefficient $\gamma$, and average polymer density $\varphi$
    in the unperturbed globule  as function of $\chi$ obtained using numerical SCF results (symbols) and using 
    analytic dependences (\ref{eq:gamma1_tau}), (\ref{eq:gamma_chi}), and(\ref{eq:chi_tau}), respectively
    (solid lines).}
  \label{fig:mugammaphi}
\end{figure}

\subsubsection{Weakly extended globule}

Let us assume that upon extension the globule 
undergoes only the shape deformation whereas its volume is conserved: The sphere transforms into a prolate uniaxial ellipsoid (often called spheroid), 
the major axis of the ellipsoid is equal to the given end-to-end distance $D$. 
We also assume that in accordance with the earlier SCF modeling results~\cite{Polotsky:2009} in a wide range of $N$ and $\chi$ such a deformation affects only the shape of the globule, the partial parameters ($\varphi$, $\mu$, and $\gamma$) remain unchanged. 
The monomer density in the deformed globule and, therefore, its volume are assumed to be constant and controlled
by the local balance of attractive binary and repulsive ternary monomer-monomer interactions and do not change upon the globule deformation.
The shape change leads to the increase of the globule surface area in Eq.~(\ref{eq:Fglobule}) and, correspondingly, to the increase of its free energy.

The volume and the surface area of the ellipsoid are
\begin{equation} \label{eq:Vsph}
  V = \frac{\pi D b^2}{6}
\end{equation}
and
\begin{equation} \label{eq:Sglobule}
  S = \frac{\pi D b}{2}\left(\sqrt{1-\alpha^2} + \frac{\arcsin \alpha}{\alpha}\right),
\end{equation}
respectively, where $b$ is the minor axis of the ellipsoid and $\alpha = \sqrt{1-b^2/D^2}$ is the ellipticity parameter.  The asymmetry of the ellipsoid (major-to-minor axes ratio) is given by
\begin{equation} \label{eq:delta}
	\delta:=\frac{D}{b}=\left(\frac{D}{2R_0}\right)^{3/2}=x^{3/2}.
\end{equation}
where we denoted by $x$ the relative elongation of the globule with respect to its unperturbed size: 
$x:=D/(2R_0)\geq 1$. Then the ellipticity is expressed as
\begin{equation} \label{eq:ellipticity}
  \alpha = \sqrt{1-\frac{b^2}{D^2}} = \sqrt{1-\frac{1}{x^3}}.
\end{equation}

Inserting Eqs~(\ref{eq:Sglobule})-(\ref{eq:ellipticity}) into Eq. (\ref{eq:Fglobule}) we obtain the free energy of the prolate ellipsoidal globule:
\begin{equation} \label{eq:Fglobule:2}
	F_{globule} = \mu N + \gamma S = \mu N + \gamma S_0\,  g(x) , 
\end{equation}
where
\begin{equation} \label{eq:g}
  g(x)=\frac{S}{4\pi R_0^2} = \frac{1}{2x} + \frac{1}{2} \frac{x^2}{\sqrt{x^3-1}}\arcsin\sqrt\frac{x^3-1}{x^3}
\end{equation}
Monomer chemical potential in extended ellipsoidal globule $\mu_{globule}$ and the reaction force $f_{globule}$ are defined as
\begin{equation} 
	\mu_{globule} = \frac{\partial F_{globule}}{\partial N} = \mu + 
	\gamma \left[\frac{dS_0}{dN}\, g(x) + S_0\, g'(x)\, \frac{dx}{dN}\right]
\end{equation}
and
\begin{equation} 
	f_{globule} = \frac{\partial F_{globule}}{\partial D} =	\gamma S_0\, g'(x)\, \frac{dx}{dD} = 2\pi R_0 \gamma\, g'(x), 
\end{equation}
respectively. Hence, the restoring force is proportional to the derivative of $g(x)$ which is equal to
\begin{equation} \label{eq:g_prime}
    g^\prime(x) = \frac{1}{2} \left[-\frac{1}{x^2} + \frac{3x}{2(x^3-1)} +
    \frac{x(x^3-4)}{2(x^3-1)^{3/2}} \arcsin\sqrt\frac{x^3-1}{x^3}\right]
\end{equation}

\begin{figure}[t] 
  \begin{center}
    \includegraphics[width=7cm]{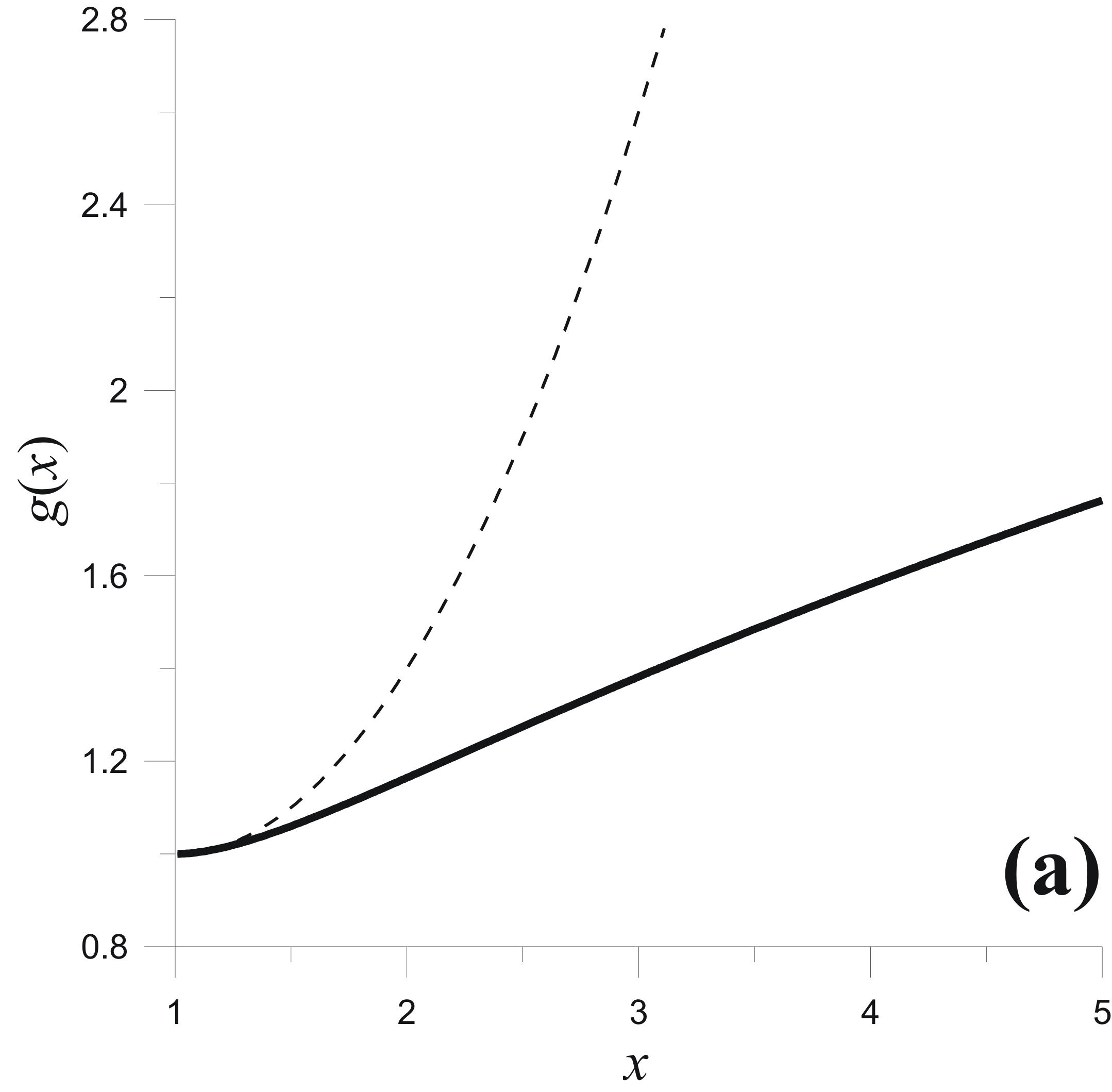}
    \includegraphics[width=7cm]{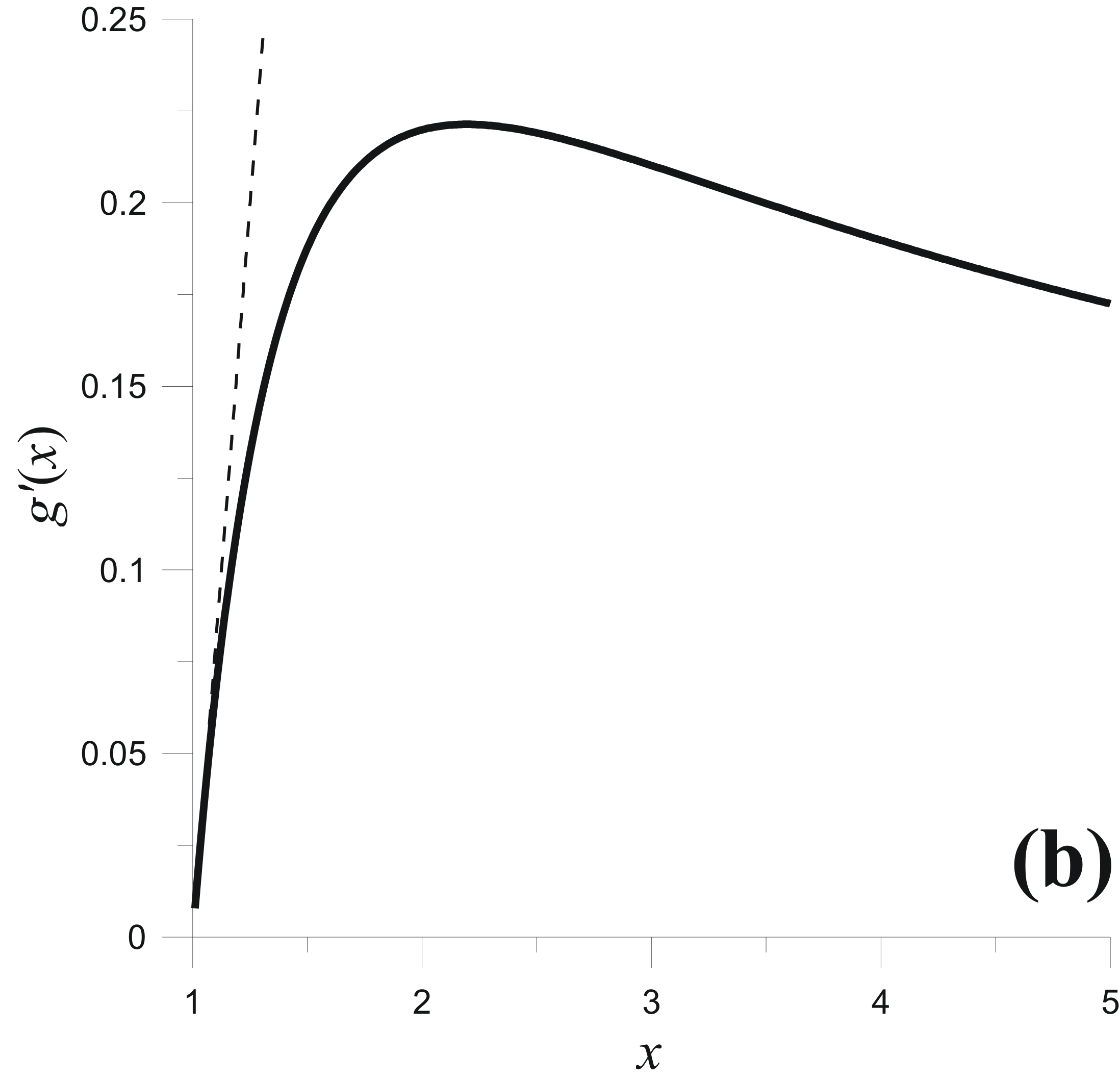}
  \end{center}
  \caption{Relative increase of the ellipsoid surface area $g = S/S_0$ as a function of the relative elongation $x=D/(2R_0)$
  (a) and its derivative (b). Solid lines - exact result, dashed lines - approximations (\ref{eq:gg_appr}).}
  \label{fig:gg_prime}
\end{figure}

Figure \ref{fig:gg_prime} shows the plots of $g(x)$ and its derivative $g'(x)$. 

At small extensions
\begin{equation}  \label{eq:gg_appr}
  \left.\begin{aligned}
    g(x)  & \simeq 1 + \frac{2}{5}(x-1)^2 \, \\
    g'(x)  & \simeq \frac{4}{5}(x-1) \,
  \end{aligned} \right. \, , \quad x-1 << 1
\end{equation}
i.e. the deformation bears elastic character, and the free energy change and the corresponding reaction force depend on the absolute value of the deformation
\begin{equation}  \label{eq:F_globule_appr}
  F_{globule}  \simeq \mu N + \gamma \left[ S_0 + \frac{2\pi}{5} (D-2R_0)^2 \right] \, ,
\end{equation}
\begin{equation}  \label{eq:force_globule_appr}
  f_{globule}  \simeq \frac{4\pi}{5}\, \gamma (D-2R_0) \, ,
\end{equation}
\begin{equation}  \label{eq:mu_globule_appr}
  \mu_{globule} \simeq 
  \mu + \frac{8\pi\gamma R_0^2}{3N} - \frac{8\pi\gamma}{15} (D-2R_0)\cdot\frac{R_0}{N} .
\end{equation}

The function $g'(x)$ passes through a maximum at $x\approx 2.19416$ and then decreases with extension. 
Asymptotic dependences at $x\gg 1$ are $g(x)\simeq \frac{\pi\sqrt{x}}{4}$ 
and $g'(x) \simeq \frac{\pi}{2\sqrt{x}}$. 
The force decrease with extension indicates an absolute instability of 
the system in this range of extension. This behavior reflects the well known Rayleigh instability~\cite{Rayleigh:1882} of a liquid droplet under uniform uniaxial extension. 
A connectivity of the monomer units within a polymer chain suggests a \emph{microphase segregation} within 
the extended globule. 
In addition to the compact globular phase, 
the second microphase should correspond to another deformation regime, namely, stretched conformation 
of the chain ($D\gg 2R_0$). 

All the formulae derived  above are equally applied to the globular microphase in the two-phase regime. 
Such a "depleted" globule loses certain fraction of its monomer units which are transferred 
to the extended phase, 
and the number of monomer units in the globule is $n < N$ and its extension is equal to $d<D$. 
The values of $n$ and $d$ are determined by the phase equilibrium conditions, 
To use the expressions obtained in this section in the range of extension corresponding 
to microphase segregated globule one should simply change $N$ to $n$ and $D$ to $d$. 
All partial characteristics ($\varphi$, $\mu$, and $\gamma$) depend neither on the number of 
segments in the globular phase, $n$
nor on the deformation $d$.

\subsection{Strongly stretched chain (unfolded globule)}\label{subsec:chain}
Now consider the limit of large $D$, $D\gg 2R_0$, where the globule is completely unfolded and 
the chain is strongly stretched,
Figure \ref{fig:globules}~c.

Let us assume that the probability of a contact between different monomer units 
(which are not nearest neighbors in the chain) is negligibly small. 
This permits to use for theoretical analysis the model of ideal chain. The first theory of stretching of the ideal polymer chain was 
elaborated more than 50 years 
ago~\cite{Volkenshtein:1955, Birshtein:1958}. 
The theory was developed for the constant force ensemble; correspondingly, 
the Gibbs free energy was calculated. 
A merit of this approach was the automatic account of finite chain extensibility which 
allows one to go beyond the range of applicability of Gaussian approximation for the chain elasticity.


Let us take advantage of this approach and ``switch'' temporarily from the fixed stretching ensemble 
studied in the present paper to the fixed force ensemble. 
Using the principle of equivalence of thermodynamic ensembles we shall interpret 
the extension vs. force dependence $D=D(f)$ as an inverse form of the force-extension dependence $f = f(D)$ 
and switch from the Gibbs free energy $G_{chain}$ to the Helmholz free energy $F_{chain}$. 

\begin{figure}[t] 
  \begin{center}
    \includegraphics[width=7cm]{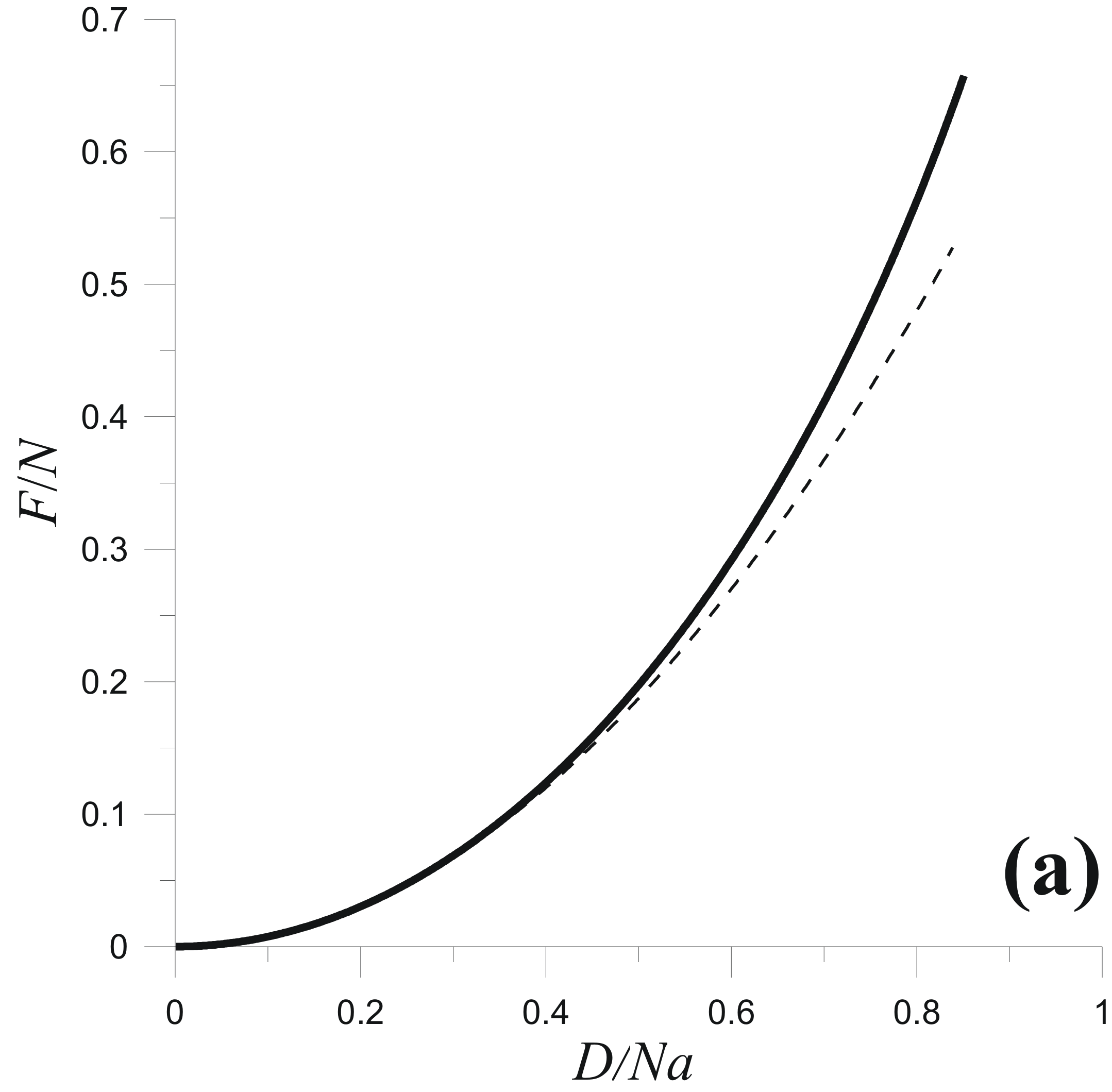}
    \includegraphics[width=7cm]{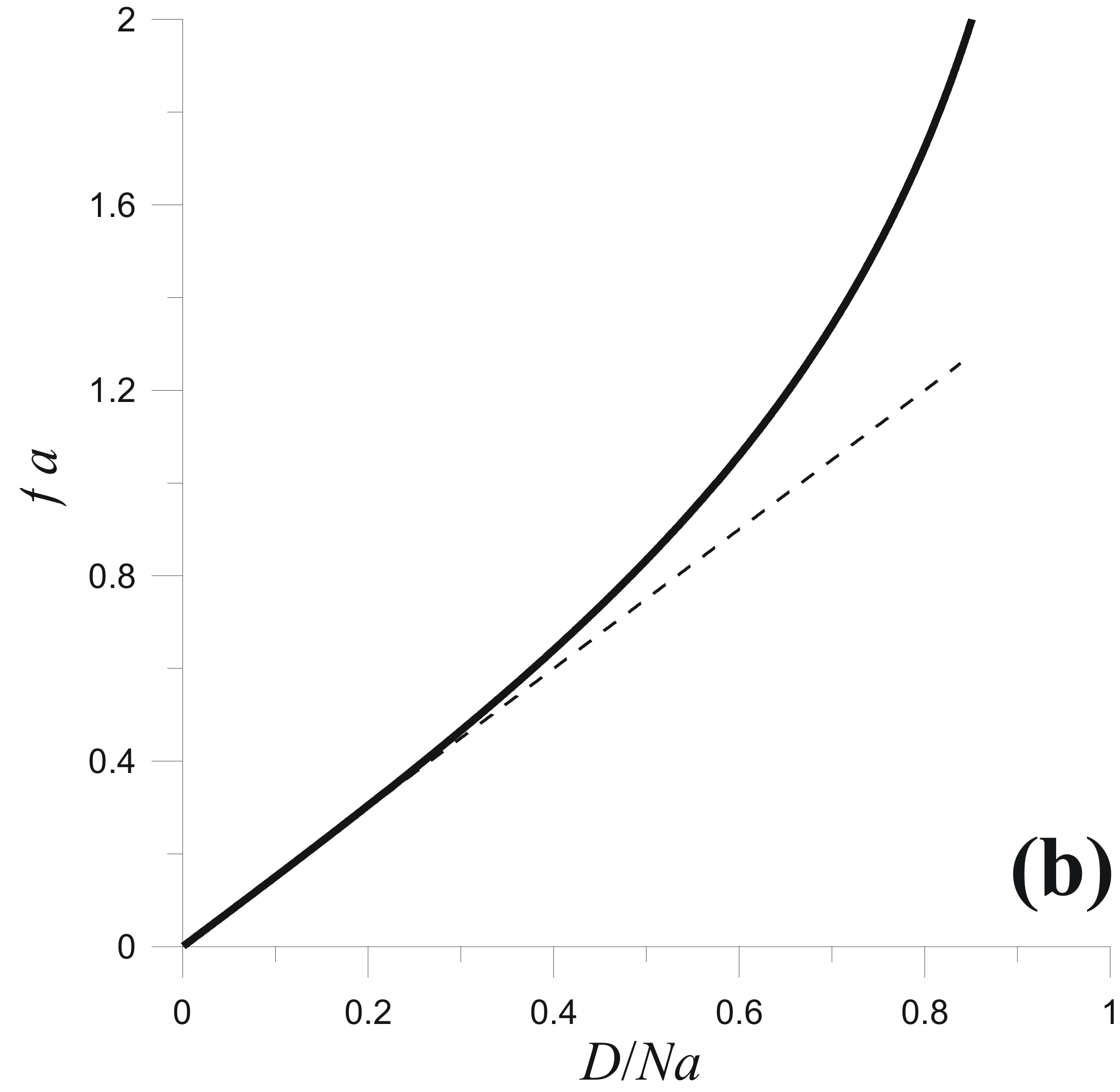}
  \end{center}
  \caption{Helmholz free energy (a) and restoring force (b) of the open chain conformation as a function of the relative
  extension. Solid line - freely-joined chain model, dashed line - approximations (\ref{eq:forcechain_appr}).}
  \label{fig:ff_chain}
\end{figure}

For the following comparison with the SCF results~\cite{Polotsky:2009} we model the chain as a random walk on a cylindrical lattice under the action of a force $f$ directed along the $z$-axis. 
Let the probability to make a step either in $r$ or in $z$ direction is given by $\lambda_1$, 
the probability of a step in ``$rz$''-direction, i.e. simultaneously changing both $r$ and $z$ 
coordinates by $\pm 1$ is $\lambda_2$ whereas the rest, $\lambda_0=1-4\lambda_1-4\lambda_2$ 
is the probability to change only the angular coordinate $\phi$. 
Then the partition function of a monomer unit is given by
\begin{equation}  \label{eq:wchain_f}
  \begin{split}
    w  & = (\lambda_0 + 2\lambda_1) e^0 + (\lambda_1 + 2\lambda_2) e^{f a} + (\lambda_1 + 2\lambda_2) 
    e^{-f a} \\ 
    & = 1-2\lambda_1 - 4\lambda_2 + 2(\lambda_1 + 2\lambda_2)\cosh\left(f a\right)
    = 1 + \frac{1}{2k} \left[\cosh\left(f a\right) - 1\right]
  \end{split}
\end{equation}
where $k := 1/(4\lambda_1 + 8\lambda_2)$
Then the partition function of the chain comprising $N$ monomer units is
\begin{equation}  \label{eq:Zchain_f}
  Z  = \left\{1 + \frac{1}{2k} \left[\cosh\left(f a\right) - 1\right]\right\}^N
\end{equation}
The logarithm of the partition function gives us \emph{the Gibbs free energy}, 
\begin{equation}  \label{eq:Gchain_f}
  G_{chain} = - \log Z  = -N\cdot \log \left\{1 + \frac{1}{2k} \left[\cosh\left(f a\right) - 1\right]\right\}
\end{equation}
Once the partition function and the Gibbs free energy are known, the average chain extension $D$ 
corresponding to the applied force $f$ can be immediately found:
\begin{equation}  \label{eq:Dchain_f}
  D = - \frac{\partial G_{chain}}{\partial f }  = 
  Na\cdot \frac{\sinh\left(f a\right)}{2k + \cosh\left(f a\right) - 1}
\end{equation}
Eq.~(\ref{eq:Dchain_f}) can be considered as an implicit dependence of the force $f$ on the extension $D$, Figure \ref{fig:ff_chain}.

The monomer chemical potential follows from the Gibbs free energy
\begin{equation}  \label{eq:muchain_f}
  \mu_{chain} = G_{chain}/N = -\log \left\{1 + \frac{1}{2k} \left[\cosh\left(f a\right) - 1\right]\right\}
\end{equation}

In order to return to the fixed extension ensemble, use the following relation between Gibbs and Helmholz 
free energies (i.e. perform the Legendre transform):
\begin{equation} \label{eq:Fchain}
  F_{chain} = G_{chain} + D\cdot f
\end{equation}
The free energy and the chemical potential in Eqs~(\ref{eq:Gchain_f}), (\ref{eq:muchain_f}), 
and (\ref{eq:Fchain}) can be expressed as a function of $D$  using the relation between $f$ and $D$
given by Eq.~(\ref{eq:Dchain_f}).

A simple form of thermodynamic functions' dependences on the extension $D$ can be obtained at $fa \ll 1$, i.e. in the limit of weak deformations. 
Then from Eq.~(\ref{eq:Dchain_f}) we find
\begin{equation}  \label{eq:Dchain_f_appr}
  D = \frac{Na}{2k} \cdot fa  
\end{equation}
from what follows
\begin{equation}  \label{eq:forcechain_appr}
  f_{chain} \simeq 2 k \cdot \frac{D}{Na^2},
\end{equation}
\begin{equation}  \label{eq:muchain_appr}
  \mu_{chain} \simeq -k \cdot \frac{D^2}{N^2 a^2},
\end{equation}
\begin{equation} \label{eq:Fchain_appr}
	F_{chain} \simeq k \cdot \frac{D^2}{Na^2}
\end{equation}
i.e the elastic free energy has a Gaussian form. 

Remarkably all the expression derived in this subsection contain a single parameter $k$ 
which defines the conditions of a random lattice walk for the free chain.

Note that with an increase in the degree of extension $D/Na$, 
the growth of both $F$ and $f$ calculated by exact formulae, Eqs (\ref{eq:Fchain}) and (\ref{eq:Dchain_f}), 
is more pronounced than that in the corresponding approximate expressions, Eqs (\ref{eq:Fchain_appr}) 
and (\ref{eq:forcechain_appr}). This is a manifestation of finite chain extensibility. 
At maximum extension, $D/Na\to 1$, the reaction force tends to infinity as 
$fa \sim -\log\left(1-\frac{D}{Na}\right)$, see  Figure \ref{fig:ff_chain}.

In the two-phase regime, the part of the chain consisting of $(N-n)$ monomer units is stretched and 
coexists in equilibrium with the globular phase containing $n$ monomer units. 
The expressions obtained in this subsection can be used for description of the stretched chain segment in the 
two-phase regime provided 
$N$ is replaced by $N-n$ and $D$ is replaced by $D-d$. 
The values of $n$ and $d$ at given $D$, $N$, and $\chi$ are determined from the equilibrium conditions 
for the two-phase system, see Eq.~(\ref{eq:eqconds}) below.

\subsection{Tadpole conformation}\label{subsec:tadpole}
In order to describe a microphase-segregated tadpole conformation 
with coexisting globular head and stretched tail, Figure \ref{fig:globules}~b, 
we assume that the globular head comprises $n$ monomers and has a shape of prolate ellipsoid with 
the major axis length equal to $d$. 
Correspondingly, the tail consists of $N-n$ monomers and stretched at the distance $D-d$. 
Then the free energy of the tadpole conformation can be written as
\begin{equation}  \label{eq:Ftdp}
    F_{tadpole} = F_{globule}(n, d) + F_{chain}(N-n, D-d)
\end{equation}

Note that we consider coexistence of only \emph{one} globule with stretched tail(s) 
``attached'' to the globule at one or at both sides, respectively. 
This minimizes the globule surface at given number of monomer units $n$ in the globular head.

The tadpole free energy $F_{tadpole}$ in Eq.~(\ref{eq:Ftdp})
should be minimized with respect to $n$ and $d$ at each given value of $D$. 
It is straightforward to show that the minimization conditions are  equivalent 
to equality of chemical potentials 
and reaction forces 
in globular and stretched phases, respectively
\begin{equation}  \label{eq:eqconds}
  \left\{\begin{aligned}
    \mu_{globule}(n, d) & = \mu_{chain}(N-n, D-d) \\
    f_{globule}(n, d) & = f_{chain}(N-n, D-d).
  \end{aligned} \right.
\end{equation}

Phase equilibrium conditions, Eq.~(\ref{eq:eqconds}) allow obtaining 
all the equilibrium characteristics of the system in the two-phase regime. 
Inserting the equilibrium values of $n$ and $d$ into Eq.~(\ref{eq:Ftdp}), 
gives the equilibrium free energy of the tadpole. 
Comparison of the latter with the free energies of one-phase states, i.e. with $F_{globule}$ and $F_{chain}$, 
enables us to find the boundary values of extension $D_1$ and $D_2$, 
which correspond to the transitions from the elongated globule to the tadpole and from the tadpole to the stretched chain conformation, 
respectively.
The range of extensions $D_1\leq D\leq D_2$ corresponds to the
thermodynamic stability of the tadpole conformation.

\section{Numerical results} \label{sec:results}

In the general case, complete exact solution for the model introduced above cannot be obtained analytically, hence, in the present section, we perform a numerical analysis of our system and compare the results with those obtained in Ref.~\cite{Polotsky:2009} using SF-SCF approach.

The numerical values of the partial parameters $\varphi$, $\mu$, and $\gamma$ are found from independent SCF modeling of free globules, see Appendix and Table 1.  $\varphi$ and $\mu$ can be also calculated using the Flory theory, see Eqs~(\ref{eq:chi_tau}) and (\ref{eq:gamma1_tau}), where the dependences are given in terms of polymer volume fraction $v$ related to the concentration as $v\equiv\varphi a^3$. 

The numerical value of the parameter $k$ that defines the law of lattice walk and, correspondingly, the elasticity of the open chain, is also chosen in the way that enables us direct comparison of theoretical and SCF modeling results. 
Namely, the nearest-neighbor and next-to-nearest neighbor step probabilities $\lambda_1$ and $\lambda_2$, respectively, were set in~\cite{Polotsky:2009} as follows: $\lambda_1=\lambda_2=1/9$. This gives 
$k=1/(4\lambda_1 + 8\lambda_2)=3/4$, 
and it is this value that is used in the numerical calculations whereas in analytical formulae $k$ will be kept as an independent parameter.

\subsection{Force-extension curves and comparison with SCF results}

\begin{figure}[t] 
  \begin{center}
    \includegraphics[width=10cm]{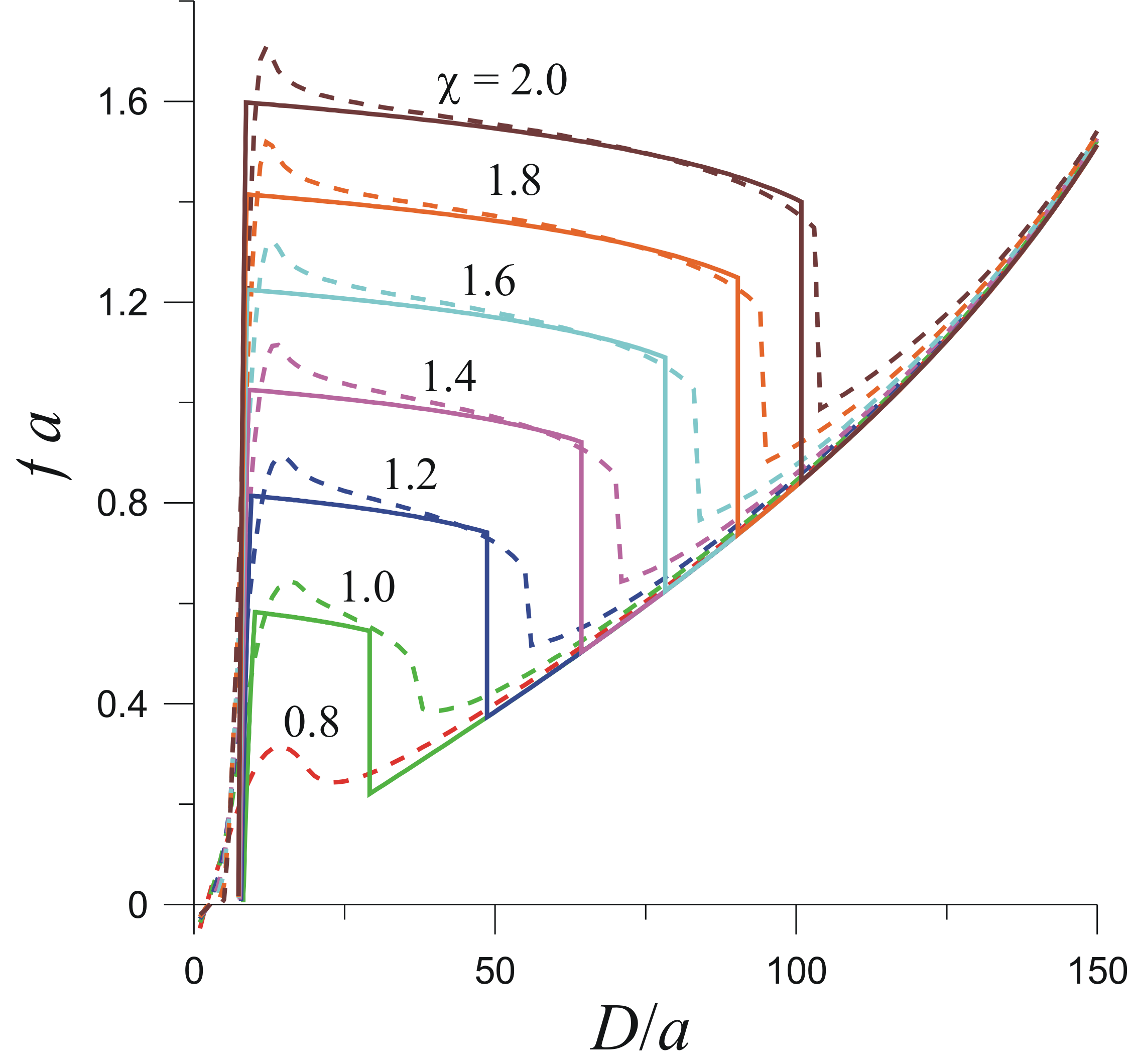}
  \end{center}
  \caption{Force extension curves for the globule with $N=200$ at various values of $\chi$. Solid lines - theory, 
  dashed lines - SCF modeling.}
  \label{fig:force_200}
\end{figure}

\begin{figure}[t] 
  \begin{center}
    \includegraphics[width=10cm]{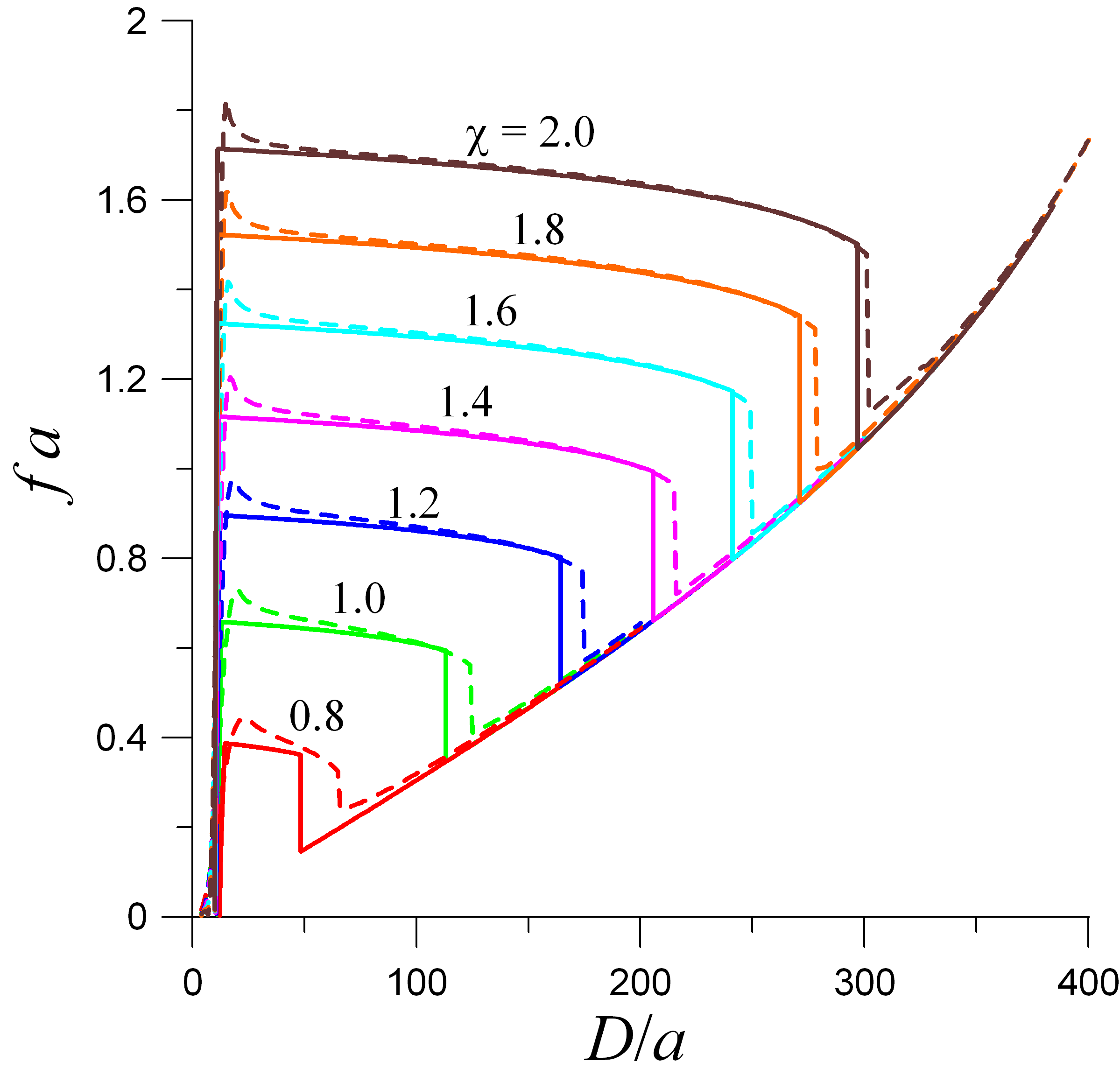}
  \end{center}
  \caption{Force extension curves for the globule with $N=500$ at various values of $\chi$. Solid lines - theory, 
  dashed lines - SCF modeling.}
  \label{fig:force_500}
\end{figure}

Force-extension curves 
calculated using Eqs~(16)-(20), (\ref{eq:Gchain_f})-(\ref{eq:Fchain}), (35), and (36)
are shown in Figures \ref{fig:force_200} and \ref{fig:force_500} together with those obtained in ref.~\cite{Polotsky:2009} 
using SCF numerical approach.

As it can be seen from Figures \ref{fig:force_200} and \ref{fig:force_500}, the theoretical curves reproduce very well the 
shape of the force-extension curves obtained in SCF modeling. 
In both cases, force growth in the ranges of small and large $D$ 
and a weakly decreasing quasi-plateau in the intermediate $D$ range 
terminated by a force drop are observed. 

Moreover, the theory parameterization based on the SCF calculation of partial parameters of the {\it free globule} as a function
of $\chi$, Appendix~\ref{app:parameters},  
leads to a very good quantitative agreement between the theory and the SCF modeling. 


Remarkably there are small divergences between the theoretical and the SCF modeling curves 
in the vicinity of the transition points between one- and two-phase state 
(globule-tadpole and tadpole - open chain, respectively).
We anticipate that these divergences are related to the assumption of the constant density of the globular phase used in the theoretical 
model.
We will return to this model applicability issue later.

\subsection{Phase diagram} \label{subsec:results:phd}
\begin{figure}[t] 
  \begin{center}
    \includegraphics[width=10cm]{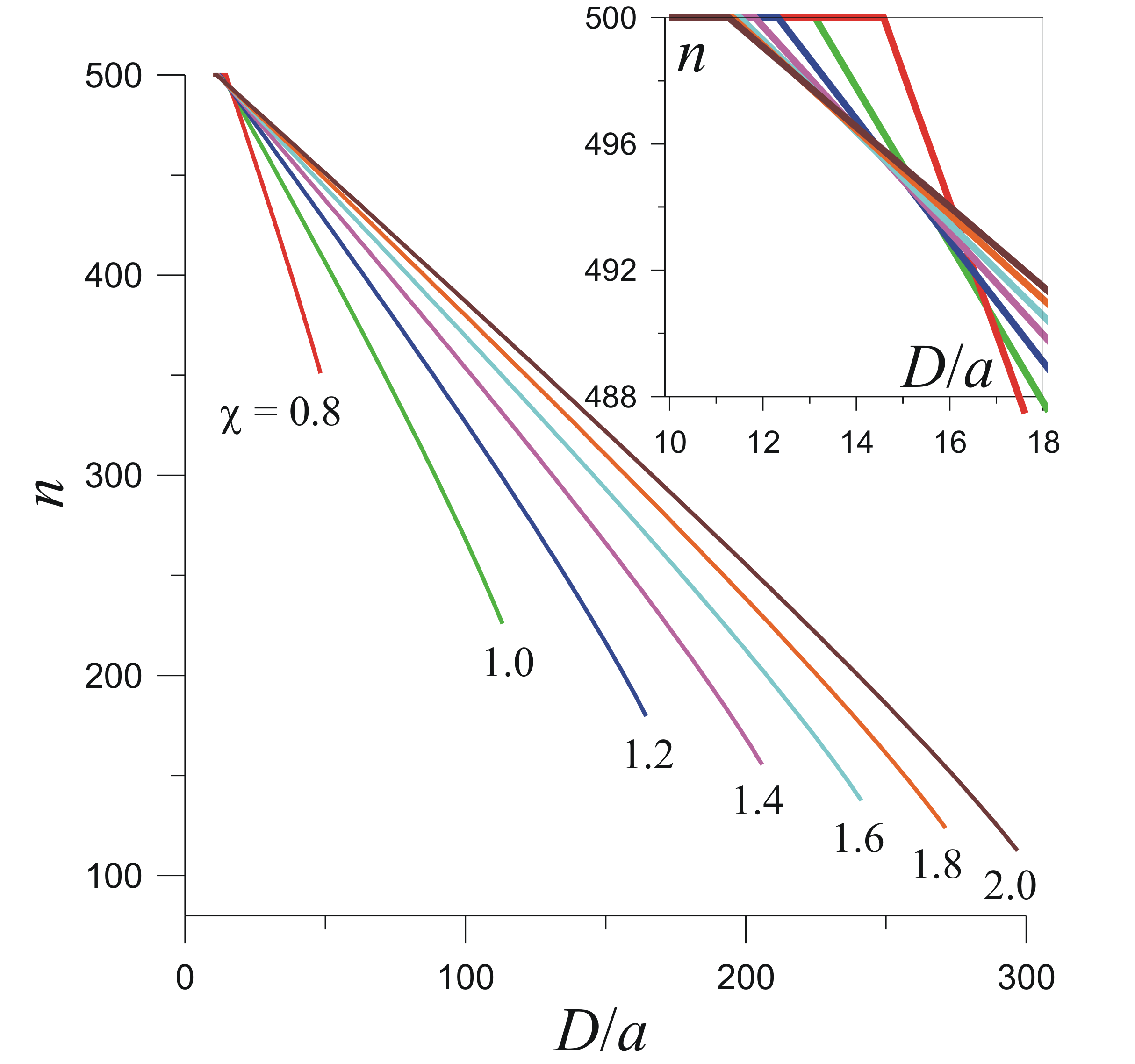}
  \end{center}
  \caption{Number of monomer units in the globular phase in the extended globule with $N=500$ at various values of $\chi$.}
  \label{fig:Ngl_500}
\end{figure}

\begin{figure}[t] 
  \begin{center}
    \includegraphics[width=10cm]{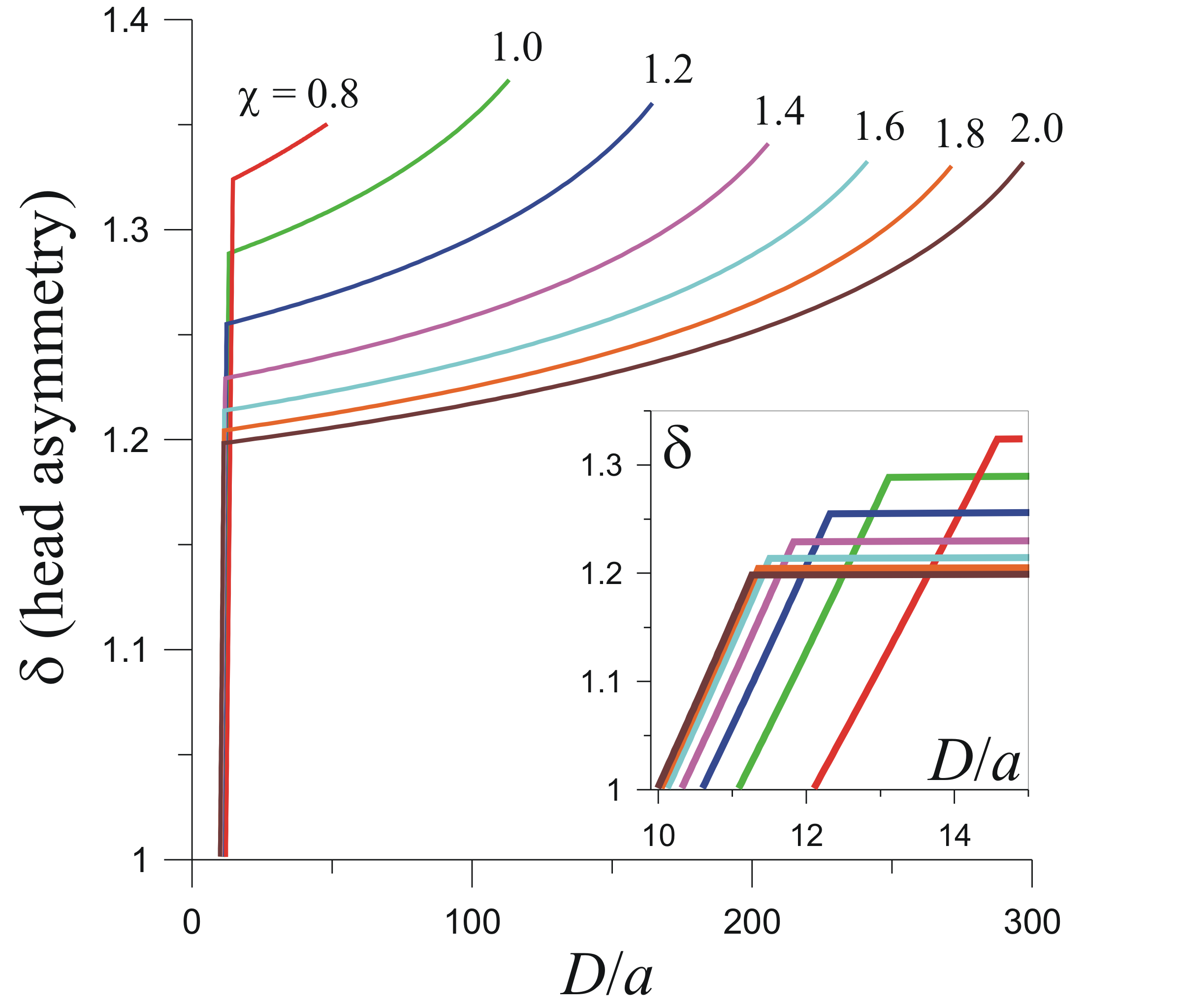}
  \end{center}
  \caption{Asymmetry ratio of the globular phase in the extended globule with $N=500$ at various values of $\chi$.}
  \label{fig:asymmetry_500}
\end{figure}

The good quantitative agreement between the force-extension curves obtained by means of the SCF modeling and the theoretical ones
allows us to use it for studying other properties of the stretched globule that were only qualitatively studied in~\cite{Polotsky:2009}. 
Figures \ref{fig:Ngl_500} and \ref{fig:asymmetry_500} show typical deformation dependences of the number of monomer units, $n$, 
in the globular head and its asymmetry, $\delta$, respectively. 
One can see a progressive increase in the asymetry reflecting the extension of the globule 
in the one-phase regime. 
Then, in the two-phase regime, the number of monomer units, $n$, in the globular phase 
starts to decrease approximately linearly with $D$, 
at the same time the asymmetry $\delta$ of the tadpoles head increases but its magnitude remains not too large. 
When the two-phase region comes abruptly to the end, the number of monomer units in the head $n$ is still large. 
After that, as $D$ further increases, the globular phase does not exist anymore.

\begin{figure}[t] 
  \begin{center}
    \includegraphics[width=10cm]{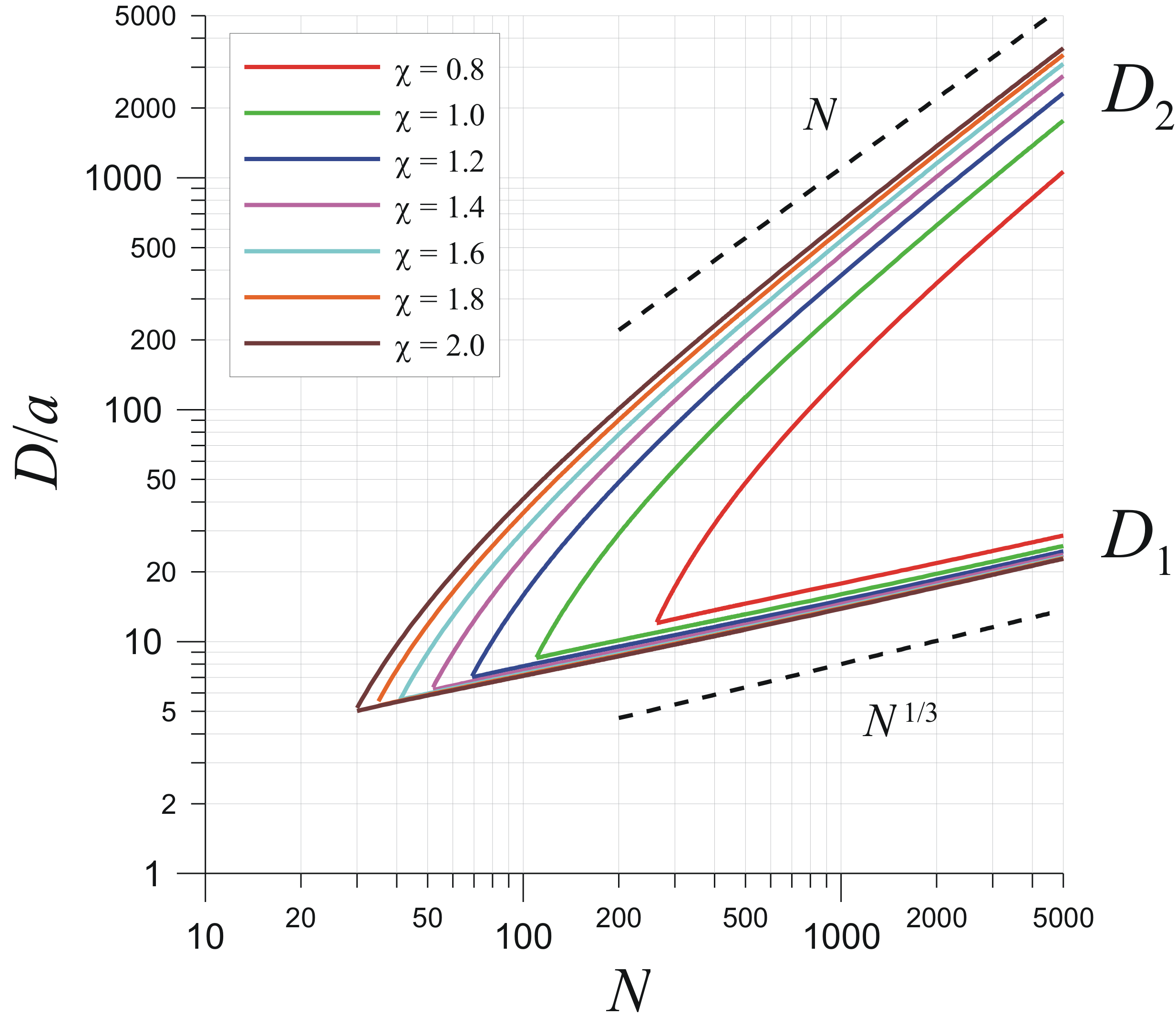}
  \end{center}
  \caption{Phase diagram of unfolded globule (limits of stability for the tadpole structure; first and second transition 
  points $D_1$ and $D_2$) at various values of $\chi$. Dashed lines show scaling according to Eqs~(\ref{eq:tp1:D}) and
  (\ref{eq:tp2:D}).}
  \label{fig:phd_D}
\end{figure}

The obtained dependences of $D_1$ and $D_2$ on $N$ and $\chi$ 
allow us to construct a phase diagram of the system, Figure \ref{fig:phd_D}. 
With a decrease in the chain length, the range of stability for the tadpole conformation 
narrows and $D_1(N), D_2(N)$ curves for two transitions  meet at a certain point $(N_{cr}, D_{cr})$. This is a \emph{critical point} for our system and in this point, i.e. for certain chain length $N=N_{cr}$, the range of extensions where the tadpole conformation corresponds to the global free energy minimum degenerates into a single point At this particular extension, $D=D_{cr}$, the ellipsoid, the tadpole, and the open chain conformation have the same free energy, thus coexisting in equilibrium. 

The value $N_{cr}$ also implies the minimal chain length at which microphase segregation within the extended globule can occur. 
$N_{cr}$ depends on the interaction parameter $\chi$: it is smaller for more solvophobic polymer. 

\begin{figure}[t] 
  \begin{center}
    \includegraphics[width=10cm]{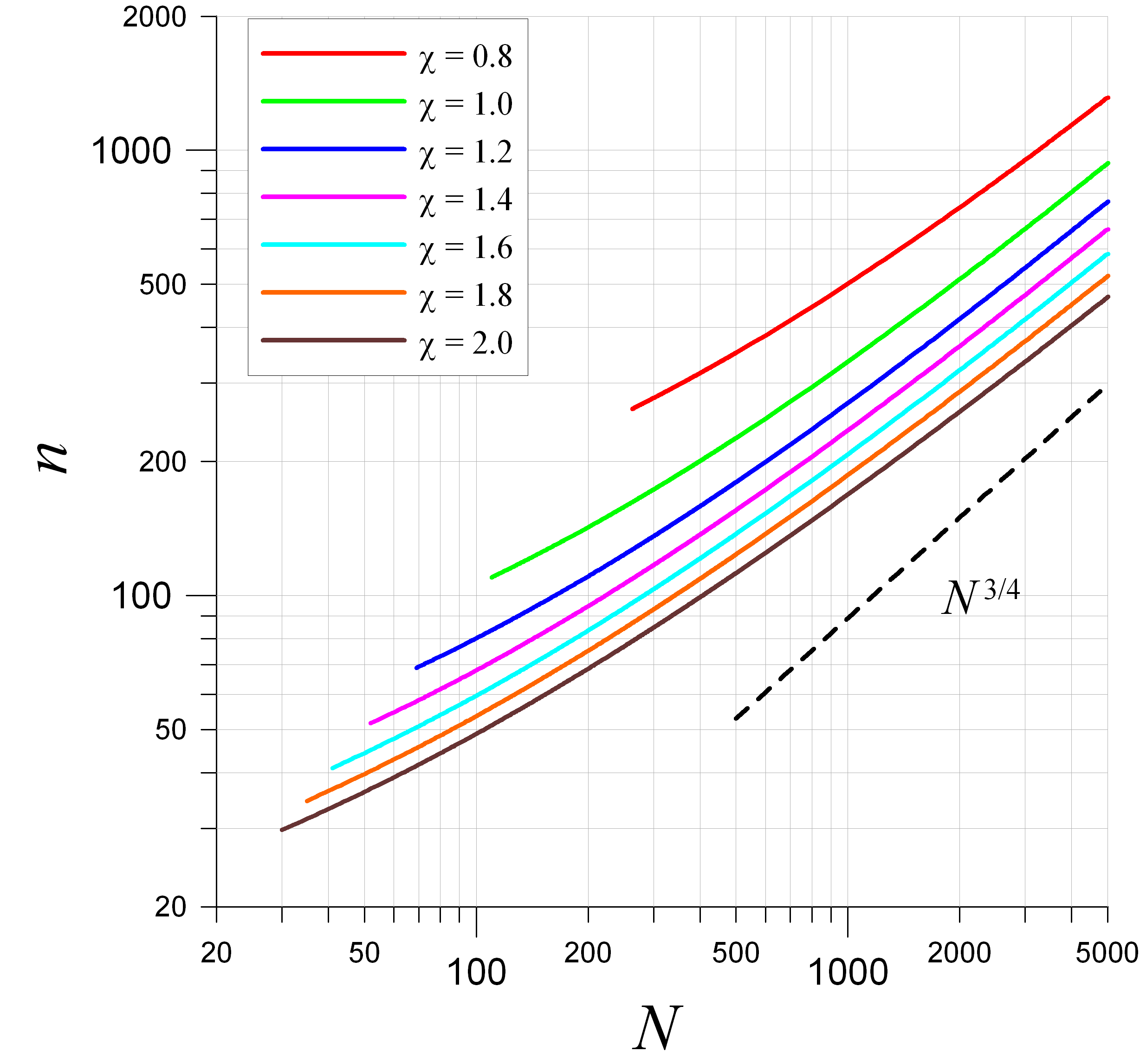}
  \end{center}
  \caption{Number of monomers in the globular head of the tadpole in the unravelling transition point at various values of
  $\chi$. Dashed line shows scaling according to Eq.~(\ref{eq:tp2_n_appr}).}
  \label{fig:phd_Ngl}
\end{figure}

\begin{figure}[t] 
  \begin{center}
    \includegraphics[width=10cm]{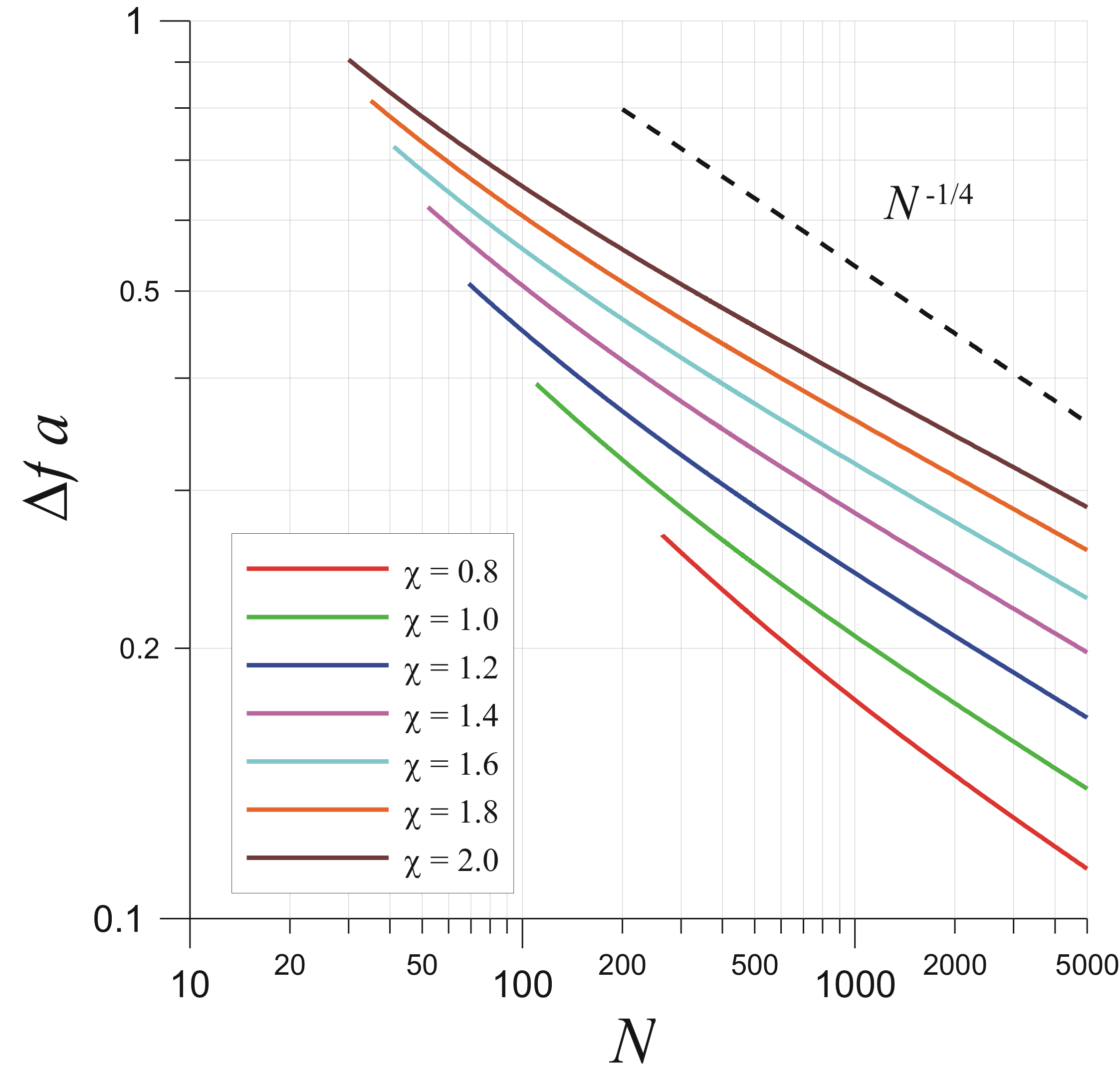}
  \end{center}
  \caption{Jump in the reaction force (restoring force) in the unravelling transition point at various values of $\chi$. 
   Dashed line shows scaling according to Eq.~(\ref{eq:tp2_deltaf}).}
  \label{fig:phd_delta_force}
\end{figure}

\begin{figure}[t] 
  \begin{center}
    \includegraphics[width=10cm]{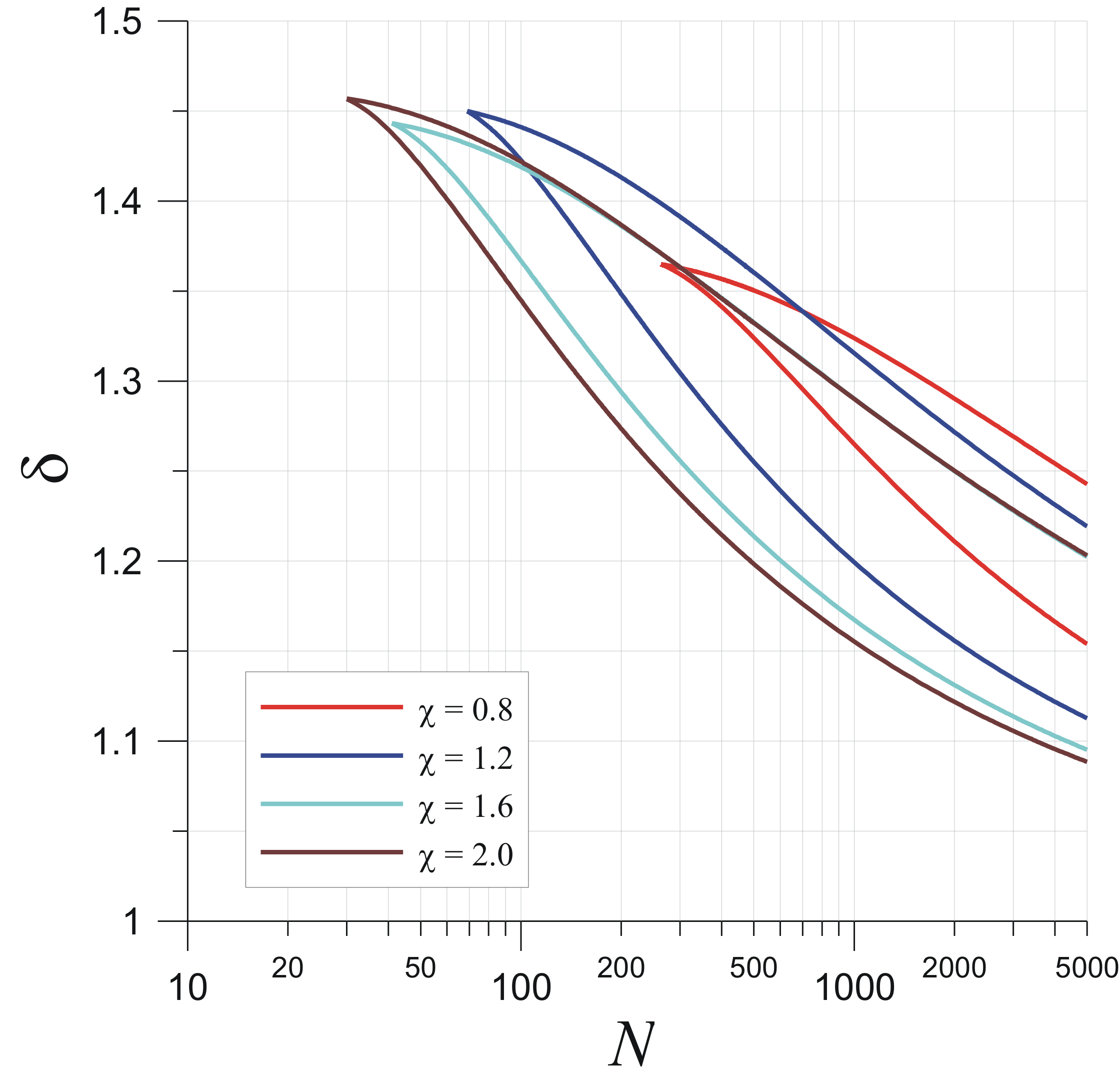}
  \end{center}
  \caption{Head asymmetry in the transition points at various values of $\chi$.}
  \label{fig:phd_asymmetry}
\end{figure}

Other properties of the system in the transition points can also be found. 
For example, Figure \ref{fig:phd_Ngl} shows the dependence of the number of monomer units, $n$, in the tadpole's head 
at the tadpole-open chain transition point 
on the overall polymerization degree $N$. 
We see that $n$ is an increasing function of $N$ and a decreasing function of  $\chi$. 
The jump in the reaction force at tadpole-open chain transition is shown in Figure \ref{fig:phd_delta_force}, its magnitude 
decays upon an increase in $N$. 
The asymmetry of the globular head of the tadpole in the transition points, Figure \ref{fig:phd_asymmetry}, 
decreases and tends to  unity as $N$ grows; in the second transition point the tadpole's head has 
obviously a more prolate shape compared to that in the first transition point.

\section{Analytical results}
In the present section we derive closed analytical expressions 
for the force-extension dependence in all deformation regimes and for the globule's properties in the transition points. 
This derivation is based on approximate expressions for thermodynamic characteristics 
of the elongated (ellipsoidall) globule and those of the stretched chain given by Eqs (\ref{eq:F_globule_appr})-(\ref{eq:mu_globule_appr})
and Eqs~(\ref{eq:forcechain_appr})-(\ref{eq:Fchain_appr}), respectively. 

In particular, we use quadratic approximation for the surface area of the prolate ellipsoida globule as a function of deformation.
The applicability range of this approximation is 
restricted by the condition $\left(\frac{D}{2R_0}\right) - 1 \equiv x-1 = \delta^{2/3} - 1 \ll 1$, see Eqs~(\ref{eq:g}), (\ref{eq:g_prime}), 
and (\ref{eq:gg_appr}) and Figure \ref{fig:gg_prime}. 
As it can be seen from Figure \ref{fig:asymmetry_500}, 
this condition are reasonably satisfied.

The Gaussian approximation for the elastic free energy of the stretched chain, Eq.~(\ref{eq:F_globule_appr})
is obtained from Eq.~(\ref{eq:Gchain_f}) in the limit of  $fa\ll 1$.  
Even though Figures \ref{fig:force_200} and \ref{fig:force_500} indicate that in the relevant range of deformations $fa\sim 1$  
we apply the approximations, (\ref{eq:forcechain_appr})-(\ref{eq:muchain_appr}) in the following calculations keeping in mind 
that it underestimates the free energy and the force (Figure \ref{fig:ff_chain}) . 

Deriving expressions below, we will also assume that both the polymerization degree $N$ and the number of monomer units in the globular phase $n$ are large, i.e. $N \geq n \gg 1$. Correspondingly, we will restrict ourselves in giving mostly the main in $N$ or $n$ term and next in $N$ or $n$ correction term in the expressions below.


\subsection{Equilibrium characteristics of the deformed globule in the two-phase regime}




In the two-phase region the free energy of a globule in the tadpole conformation is easily obtained using Eqs~(\ref{eq:Ftdp}), 
(\ref{eq:F_globule_appr}), and (\ref{eq:Fchain_appr}):
\begin{equation}  \label{eq:Ftadp_appr}
  F_{tadpole} \simeq  \mu n + 4\pi\gamma r_0^2 + \frac{2\pi\gamma}{5} (d-2r_0)^2 + k \cdot
  \frac{(D-d)^2}{(N-n)a^2}.
\end{equation}
where the first term corresponds to the volume contribution to the free energy of the tadpoles head, 
the second and the third terms describe its excess surface free energy, wherease the last term describes the
free energy of the extended tail.

Recall that $n$ is the number of monomer units in the globular head of the tadpole, 
$d$ is the head's main axis size and $r_0$ is the radius of the spherical globule containing $n$ monomer units: 
\begin{equation}
	r_0=\left(\frac{3n}{4\pi\varphi}\right)^{1/3}.
\end{equation}
By minimizing the free energy (\ref{eq:Ftadp_appr}) with respect to $n$ and $d$  (or, equivalently, applying the equilibrium conditions (\ref{eq:eqconds})), we obtain
\begin{equation}  \label{eq:eqconds_approx}
  \left\{\begin{aligned}
    & \mu + \frac{8\pi\gamma r_0^2}{3n} - \frac{8\pi\gamma}{15} (d-2r_0)\cdot\frac{r_0}{n} = 
    -k \cdot \frac{(D-d)^2}{(N-n)^2 a^2} \\
    & \frac{4\pi\gamma}{5} (d-2r_0) = 2 k \cdot \frac{D-d}{(N-n)a^2}.
  \end{aligned} \right.
\end{equation}

Let us find the long axis of the tadpole's head. By substituting $(D-d)/(N-n)$ from the second equation of the system into the first one, one arrives at a quadratic equation for $(d-2r_0)$. It gives the following solution for $d$:
\begin{equation} \label{eq:d_tadpole_appr}
	d \simeq 2\left(\frac{3n}{4\pi\varphi}\right)^{1/3} + 
	\frac{5\sqrt{k |\mu|}}{2\pi\gamma a} \left[1- \frac{\gamma}{|\mu|}\left(\frac{4\pi}{3n\varphi^2}\right)^{1/3} \right]
	\simeq 2r_0 + \frac{5\sqrt{k |\mu|}}{2\pi\gamma a}
\end{equation}
Then the tadpole's head asymmetry is given by
\begin{equation} \label{eq:asymmetry_tadpole_appr}
	\delta=\left(\frac{d}{2r_0}\right)^{3/2} \simeq \left(1 + \frac{5\sqrt{k |\mu|}}{4\pi\gamma r_0 a}\right)^{3/2}
\end{equation}
i.e. in the two-phase state the globular head is elongated, its longer axes is larger than the diameter of a 
spherical globule comprising the same number, $n$, of the monomer units.
Remarkably, the second (correction) term in Eq.~(\ref{eq:d_tadpole_appr}) is independent of $n$.
According to the $\chi$-dependences of $\gamma$ and $\mu$ this term is 
a decreasing function of $\chi$. 
A relative elongation $d/2r_0$ and related to it asymmetry parameter 
$\delta=(d/2r_0)^{3/2}$ 
of the tadpole's globular head grow with an increase in $D$ because of a 
decrease in the number of monomer units in the globular head $n$, Figure \ref{fig:asymmetry_500}. 

The $n(D)$ dependence is obtained from the second of Eq.~(\ref{eq:eqconds_approx}) and reads
\begin{equation} \label{eq:ngl_tadpole_appr}
	n \simeq N - \frac{5k (D-d)}{2\pi\gamma (d-2r_0)} \simeq N - \sqrt{\frac{k}{|\mu|}}(D-d)
\end{equation}
In the two-phase region, where $D\gg d$, the decrease of $n$ as function of $D$ 
is close to linear, the slope of this dependence is independent of $N$ and 
decreases with an increase in $\chi$ (see Table 1 in Appendix and Figure \ref{fig:Ngl_500}).

The reaction force $f$ is obtained from Eqs~(\ref{eq:eqconds_approx}) and (\ref{eq:d_tadpole_appr}) as a function of $n$ and is given by
\begin{equation} \label{eq:ftadpole_appr}
	f_{tadpole} \simeq \frac{2\sqrt{k |\mu|}}{a} \left[1- \frac{\gamma}{|\mu|}\left(\frac{4\pi}{3n\varphi^2}\right)^{1/3} \right].
\end{equation}
This equation shows that the reaction force decreases with extension: as $D$ increases, the number of monomers in the globular phase, $n$, decreases and, therefore $f$ decreases, Figures \ref{fig:force_200}, \ref{fig:force_500}.

Note that the correction term in Eq.~(\ref{eq:ftadpole_appr}) growing with a decrease in $n$ includes the interfacial tension coefficient $\gamma$ in the combination with the other partial parameters of the globule: $\gamma|\mu|^{-1} \varphi^{-2/3}$


\subsection{Transition points}
In order to find the values of deformation, $D$, corresponding to the 
boundaries (transition points) between different regimes, the free energies of possible states should be calculated and then compared;
then the boundary values of $D$ are found as intersection points of $F(D)$ dependences. 

The transitions observed in the system are those between one of the one-phase states (prolate globule or extended chain) 
and the two-phase state (tadpole conformation). 
Therefore, the system of equations for determination of the transition point includes the two-phase state equilibrium conditions, Eq. (\ref{eq:eqconds}),
and the free energy equality condition $F_{tadpole} = F_{globule}$  or  $F_{tadpole} = F_{chain}$. 
Taking into account Eq.~(\ref{eq:Ftdp}) the free energy equality condition reads
\begin{equation}  \label{eq:tp}
 \begin{split}
   & F_{globule}(n, d) + F_{chain}(N-n, D-d) = F_{globule}(N,D) \\
   \mbox{ or } \qquad & F_{globule}(n, d) + F_{chain}(N-n, D-d) = F_{chain}(N,D)
 \end{split}
\end{equation}

By solving together Eqs~(\ref{eq:eqconds}) and (\ref{eq:tp}) one obtains the threshold value of the 
 extension $D$ as well as the vlues of $n$, $d$ in the corresponding transition point.

\subsubsection{Globule - tadpole transition point} 

The critical extension for the prolate globule-tadpole transition point is obtained quite easily. 
According to our model, at this point the new stretched phase appears. This phase has an infinitely small size, hence, we should take the limit $n\to N$ and $d\to D$ in Eqs~(\ref{eq:Ftdp}) and (\ref{eq:eqconds})
This gives us the transition point
\begin{equation} \label{eq:tp1:D}
	D_1 \simeq 2\left(\frac{3N}{4\pi\varphi}\right)^{1/3} + 
	\frac{5\sqrt{k |\mu|}}{2\pi\gamma a} \left(1- \frac{\gamma}{|\mu|}\left(\frac{4\pi}{3N\varphi^2}\right)^{1/3} \right)
\end{equation}
and the corresponding reaction force in the transition point
\begin{equation} \label{eq:tp1_f}
	f_1 \simeq 
	\frac{2\sqrt{k |\mu|}}{a} \left(1- \frac{\gamma}{|\mu|}\left(\frac{4\pi}{3N\varphi^2}\right)^{1/3} \right) .
\end{equation}
The asymmetry of the ellipsoidal globule (see Eq.~(\ref{eq:delta})) in the transition point is
\begin{equation} \label{eq:tp1_delta}
	\delta_1=\left(\frac{D}{2R_0}\right)^{3/2} \simeq 
	1 + \frac{5\sqrt{k|\mu|}}{4\gamma a}\left(\frac{9\varphi}{2\pi^2 N}\right)^{1/3}.
\end{equation}

Eqs (\ref{eq:tp1:D})-(\ref{eq:tp1_delta}) describe dependences of $D_1$, $f_1$, and $\delta_1$ at the transition point 
on partial characteristics of the globule, $\varphi$, $\mu$, and $\gamma$, and on the parameter of lattice walks $k$. 
According to Eq. (\ref{eq:tp1:D}) the globule extension at the transition point,  $D_1$, scales in the main term as $D_1\sim R_0 \sim N^{1/3}$ 
at large $N$ (Figure \ref{fig:phd_D}); the absolute value of the (negative) correction term decreases as a function of $N$. 
The restoring force $f$ varies in the transition point continuously as a function of $D$, 
the leading term grows with $\chi$, the correction term scales as $N^{-1/3}$ and decays with $N$. 
The correction term in Eq.~(\ref{eq:tp1_f}) includes the already mentioned combination or the partial parameters 
$\gamma|\mu|^{-1} \varphi^{-2/3}$. 
The asymmetry of the globule in the transition point decreases as $N$ grows (Figure \ref{fig:phd_asymmetry}), 
the same is valid for the asymmetry of the globular head of the tadpole in the two-phase region.

\subsubsection{Tadpole - open chain transition point} 
Now let us consider the second transition corresponding 
to unravelling of the globular head of the tadpole.
This transition occurs, as it follows from the force-extension curves, Figures \ref{fig:force_200}, \ref{fig:force_500}, at substantial deformations. 
Assuming that $D\gg d$ and equating the free energy of the tadpole (\ref{eq:Ftadp_appr}) and that of the stratched ("open") chain (\ref{eq:Fchain_appr}) 
we obtain teh following equation:
\begin{equation}  \label{eq:tp2_cond:1}
	\mu n + 4\pi\gamma r_0^2 + \frac{2\pi\gamma}{5} (d-2r_0)^2 + k \cdot \frac{D^2 n}{N(N-n)a^2} = 0
\end{equation}
The value of $D$ can be expressed from the second equation of the system (\ref{eq:eqconds_approx}): 
\begin{equation} \label{eq:tp2:D:exact}
	D = \frac{2\pi\gamma}{5k}(N-n)a^2(d-2r_0). 
\end{equation}
Taking into account that $d-2r_0$ is given by Eq.~(\ref{eq:d_tadpole_appr}) and neglecting the correction to the head surface energy 
(the 3rd term in the lhs of Eq.~(\ref{eq:tp2_cond:1})) we get
\begin{equation}  \label{eq:tp2_cond:2}
	\mu n + 4\pi\gamma r_0^2 + 
	|\mu| n \left(1-\frac{n}{N}\right) \left[1 - \frac{\gamma}{|\mu|}\left(\frac{4\pi}{3n\varphi^2}\right)^{1/3} \right] = 0
\end{equation}
Keeping only the dominating contributions of the highest order in $n$ we obtain
\begin{equation}  \label{eq:tp2_n_appr}
   n \simeq \left(\frac{2\gamma}{|\mu|} \right)^{3/4} \left(\frac{4\pi}{3\varphi^2}\right)^{1/4} N^{3/4},
\end{equation}
i.e. the number of monomer units in the head of the tadpole at the transition point 
scales as $N^{3/4}$, see Figure \ref{fig:phd_Ngl},
similar to what has been found for a dry globule in ref. \cite{Cooke:2003}. 
The size of the tadpole's head, $d$, is easily found from Eqs~(\ref{eq:d_tadpole_appr}) and (\ref{eq:tp2_n_appr})
\begin{equation} \label{eq:tp2_dheadtdp}
	d_2 \simeq 2 \left(\frac{2\gamma}{|\mu|} \right)^{1/4} \left(\frac{3}{4\pi \varphi^2}\right)^{1/4} N^{1/4} +
	\frac{5\sqrt{k|\mu|}}{2\pi\gamma a}
\end{equation}
from what the asymmetry of the head in the transition point follows
\begin{equation} \label{eq:tp2_delta}
	\delta_2 \simeq 1 + 5\sqrt{k\varphi} \left(\frac{|\mu|}{2\gamma} \right)^{5/4}  \left(\frac{3}{4\pi}\right)^{3/4} N^{-1/4}.
\end{equation}

The extension $D_2$ at the transition point is then obtained using Eqs~(\ref{eq:tp2:D:exact}), (\ref{eq:tp2_n_appr}), and (\ref{eq:tp2_dheadtdp})
and reads:
\begin{equation} \label{eq:tp2:D}
	D_2 \simeq \sqrt{\frac{|\mu|}{k}} \cdot Na \left[1- \frac{3}{2} \left(\frac{2\gamma}{|\mu|} \right)^{3/4}
	\left(\frac{3}{4\pi \varphi^2}\right)^{1/4} N^{-1/4} \right] .
\end{equation}
The reaction force of the tadpole is obtained using Eqs~(\ref{eq:ftadpole_appr}) and (\ref{eq:tp2_n_appr}):
\begin{equation} \label{eq:tp2_ftadp}
  f_{tadpole} \simeq \frac{2\sqrt{k|\mu|}}{a} \left[1-\frac{1}{2} \left(\frac{2\gamma}{|\mu|} \right)^{3/4}
  \left(\frac{4\pi}{3\varphi^2}\right)^{1/4} 
  N^{-1/4} \right]
\end{equation}
The reaction force  of the stretched ("open") chain in the transition point is given by
\begin{equation} \label{eq:tp2_fchain}
  f_{chain}\simeq \frac{2kD}{Na^2} \simeq \frac{2\sqrt{k|\mu|}}{a} 
  \left[1-\frac{3}{2}\left(\frac{2\gamma}{|\mu|} \right)^{3/4} \left(\frac{4\pi}{3\varphi^2}\right)^{1/4} N^{-1/4} \right]
\end{equation}
Hence, the force jump in the transition point is given by
\begin{equation} \label{eq:tp2_deltaf}
  \Delta f = f_{tadpole} - f_{chain}\simeq 
  \frac{2\sqrt{k|\mu|}}{a} \left(\frac{2\gamma}{|\mu|} \right)^{3/4} 
  \left(\frac{4\pi}{3\varphi^2}\right)^{1/4} N^{-1/4}
\end{equation}
and decays as $N$ increases (Figure \ref{fig:phd_Ngl}).

\subsection{Critical point} 
Using the formulas obtained above we can also find the critical point $N_{cr}$. It can be formally estimated using the expressions (\ref{eq:tp1:D}) and (\ref{eq:tp2:D}) for $D_1$ and $D_2$ in the large globule limit 
(although  the globule consisting of $N=N_{cr}$ monomers may hardly be considered as a ``large'' one). 
\begin{equation} \label{eq:Ncr}
	N_{cr}\simeq \sqrt{\frac{6}{\pi \varphi a^3}} \left(\frac{k}{|\mu|}\right)^{3/4}
\end{equation}
In the vicinity of the theta-point all partial parameters scale as certain power law functions of relative deviation from the theta-point,
$\tau\equiv \chi/\chi_{\theta}-1$. In particular, $\varphi\sim \tau a^{-3}$, $\gamma\sim \tau^{2}a^{-2}$, $|\mu|\sim \tau^{2}$.
Hence, condition $N\gg N_{cr}$ implies $N\tau^{2}\gg 1$, that is the requirement of stability of the globular state~\cite{deGennes:1979}.

\section{Discussion}


\subsection{Statistico-mechanical analogy}
Let us start with discussion of analogy between intra-molecular conformational transition and microphase coexistence in extended polymer globule 
and well-known classical problem of gas-to-liquid transition.

\begin{figure}[t] 
  \begin{center}
    \includegraphics[width=12cm]{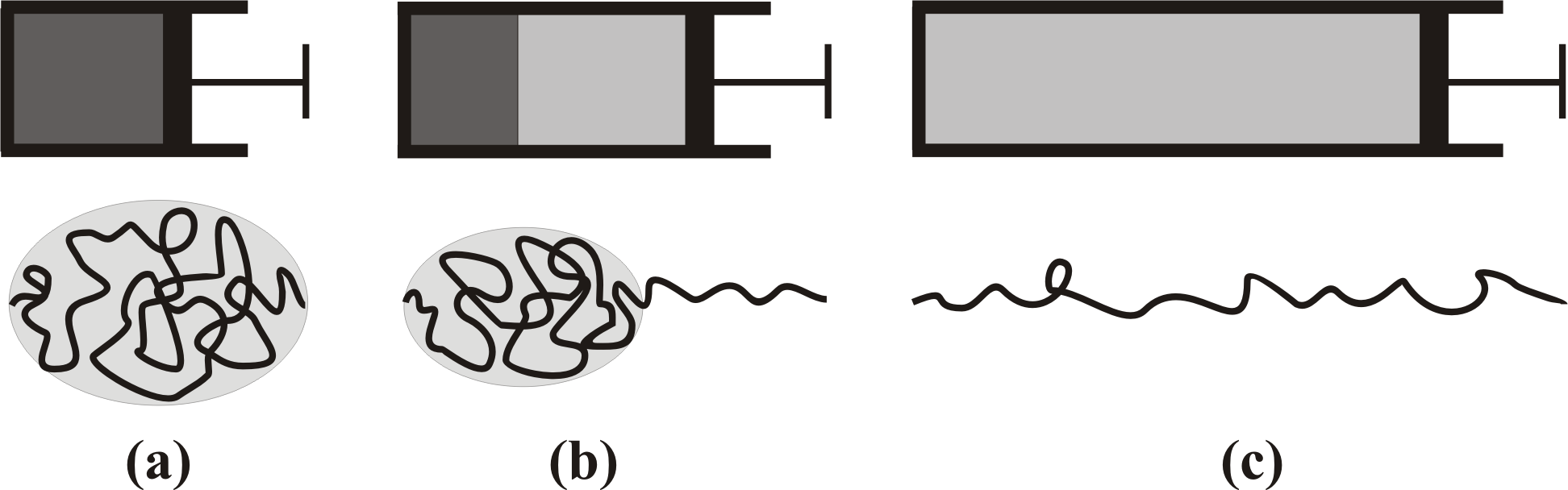}
  \end{center}
  \caption{Correspondence between states of the van der Waals gas in $(V, T)$-ensemble and conformations of stretched 
  globule in $D$-ensemble.}
  \label{fig:gas}
\end{figure}

The conformational rearrangements in the deformed globule are similar to the liquid-to-gas transition that occur 
with the van der Waals gas in the $(V, T)$-ensemble below critical temperature ($T<T_{crit}$) upon increasing the system volume $V$. 
At small $V$, the system is in the liquid state which is the counterpart of the globular state, Figure \ref{fig:gas}~a. 
An increase in volume gives rise to the vapor phase coexisting in equilibrium with the liquid one, similar microphase  coexistence 
is observed in the tadpole state of the globule, Figure \ref{fig:gas}~b. 
As $V$ further increases, the molecules move from the liquid to the vapor phase and finally, 
when $V$ is large, the system is completely in vapor (gas) state, in our system this role is played by the stretched chain state, 
Figure \ref{fig:gas}~c. 

\begin{figure}[t] 
  \begin{center}
    \includegraphics[width=14cm]{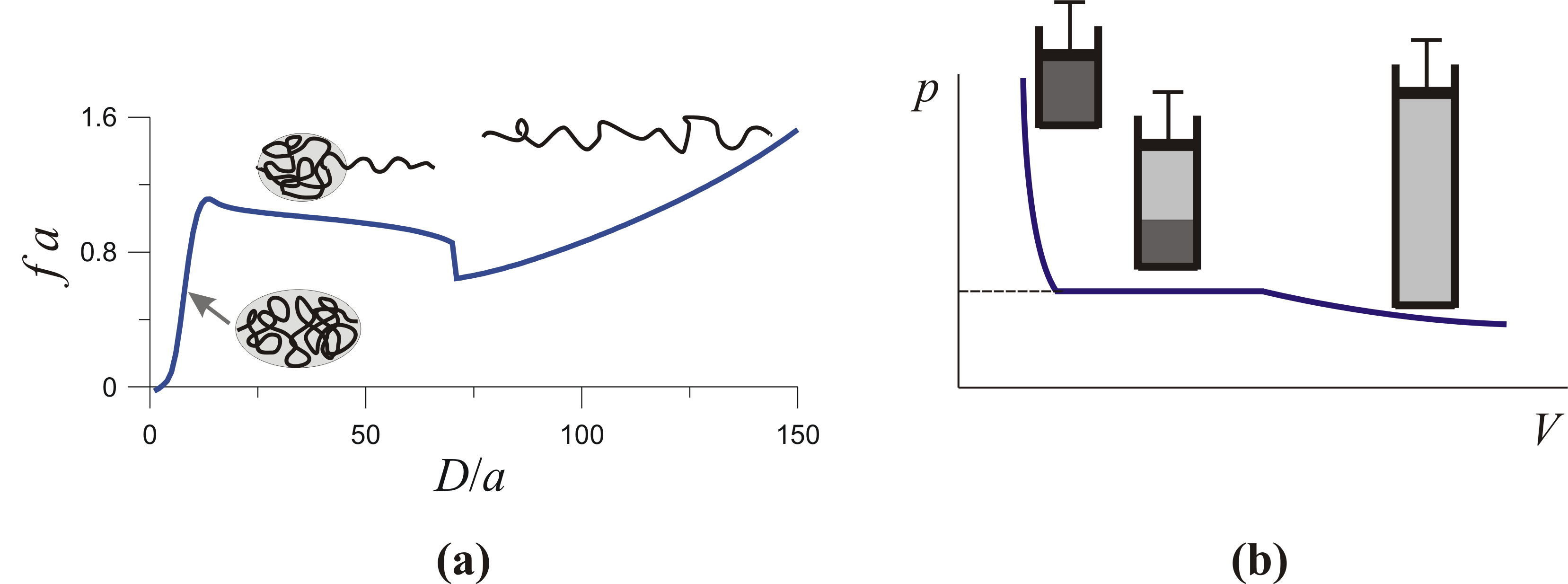}
  \end{center}
  \caption{Globule force-extension curve obtained using SCF numerical approach for $N=200$, $\chi=1.4$ in
  Ref.~\cite{Polotsky:2009} (a) and schematic $P-V$ diagram for the van der Waals gas (b).}
  \label{fig:FD_and_PV}
\end{figure}



From the point of view of conformational changes, the analogy illustrated by Figure \ref{fig:gas} is obvious. 
However, thermodynamic characteristics in our small (single polymer chain) system behave differently than those in 
macroscopic gas-liquid system. Comparison of $p=p(V)$ and $f=f(D)$ dependences, Figure \ref{fig:FD_and_PV}, indicates a marked difference. 

In both systems,  the pressure $p$ and the reaction force $f$ vary monotonically with an increase in $V$ or $D$, 
respectively,  in the one-phase states. (These changes however differ in sign because of different directions of reaction force and pressure).  
An essential difference between single polymer and macrosystem is observed in the phase coexistence regime. 
In the macrosystem [$(V, T)$- ensemble] the pressure $p$ is constant at any system volume where two phases coexist, 
Figure \ref{fig:FD_and_PV}~b. In the extended globule the reaction force dependence on $D$ is ``anomalous'':  
an increase in $D$ in the phase coexistence regime is accompanied by a decrease in $f$, Figure \ref{fig:FD_and_PV}~a. 
Although such a dependence is usually interpreted in statistical physics as a manifestation of instability, for the considered small 
(single polymer molecule) system this state cannot be avoided in the constant extension ensemble. 
In the range of $D$ corresponding to phase coexistence the decrease in the force is monotonous but at the right edge it drops abruptly. 
The system therefore demonstrates a negative extensibility (a negative elastic modulus).

\subsection{Force-extension curves and phase diagram}
Let us discuss the obtained results on equilibrium globule stretching by comparing them with
results of the SCF modeling reported in detail in~\cite{Polotsky:2009}. 
As it can be seen from Figures \ref{fig:force_200} and \ref{fig:force_500}, the quantitative theory is in a very good agreement 
with the SCF modeling results. 

Both approaches lead to the conclusion about existence of the three-state mechanism 
of extention of a homopolymer 
chain which is immersed into a poor solvent ($\chi > 1/2$) and has the globular conformation as an unperturbed state. 
Three states of the deformed globule correspond to a weakly elongated (ellipsoidal) globule and to a stretched chain at small and large $D$, respectively, 
and to coexistence of these two states within a single macromolecule at intermediate extensions.

As follows from Figures \ref{fig:phd_D} and Eq.~(\ref{eq:Ncr}), the system has a \emph{critical point}, 
i.e. there exists a minimal chain length $N_{cr}$ below which, at $N < N_{cr}$ microphase segregation in the globule is not possible. 
Here $N_{cr}$ has the order of tens of monomer units at $\chi =1.5 \div 2.0$ (the vloume fraction of the globule in this case is $\varphi a^3 \sim 1$) 
and increases to approximately 200 monomers when $\chi$ is decreased to 0.8 ($\varphi a^3 \sim 0.5$). 

The three-stage mechanism is realized provided the condition $N > N_{cr}$ is fulfilled.
On the contrary,  a small (``overcritical'') globule upon extension is deformed ``as a whole'', without intramolecular segregation. 
As follows from analytical expression (\ref{eq:Ncr}), the value of $N_{cr}$ is independent of $\gamma$, 
i.e. it is determined only by the internal characteristics of the globule, 
but includes a dependence on $k$ - the characteristic of the free chain walk,
\begin{equation*}
	N_{cr}\sim \left(\frac{k}{|\mu|}\right)^{3/4} \frac{1}{\varphi^{1/2}}
\end{equation*}

Conformation of extended globule at these overcritical conditions was studied using SCF modeling~\cite{Polotsky:2009} $[N=200, \, \chi = 0.8]$. 
It was shown that in this case the assumption about a constant polymer density $\varphi$ within the deformed globule cannot be fulfilled. 
Extension of a small overcritical globule causes not solely the shape deformation but also a volume change - 
correspondingly a decrease in $\varphi$ and depletion (or ``dissolution'') of its core. 
Emergence of this state clearly shows its thermodynamic advantage in extended small globule at weakly poor solvent. 
At the same time the value of $N_{cr}$ shows the limit of applicability of our model since the assumption about constant polymer density 
(``volume approximation'') is no more valid at small $N$, where the width of the interface (globule's interfacial layer) 
becomes comparable to the radius of the globular core. 
This is also illustrated by the deviation of $\varphi$ values found in the SCF modeling of free globules at smaller $N$ (large $1/N$) 
from the best fit on Figure \ref{fig:phi_globules} in Appendix~\ref{app:parameters}. This explains the effect observed in the SCF modeling.

\begin{figure}[t] 
  \begin{center}
    \includegraphics[width=10cm]{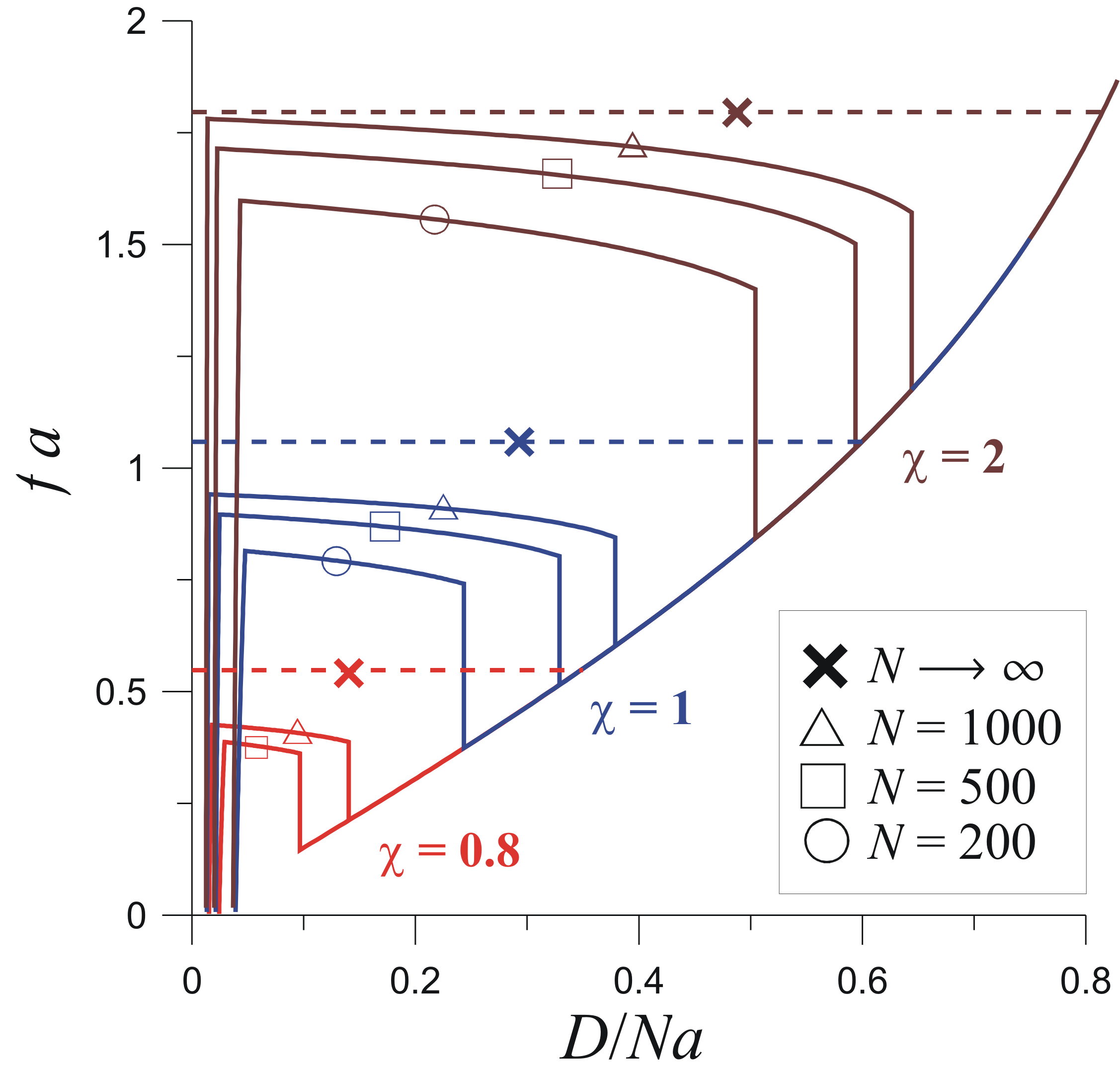}
  \end{center}
  \caption{Force vs degree of extension curves at various values of $N$ and $\chi$. Dashed lines show the position of 
  the plateau at $N\to \infty$.}
  \label{fig:force_reduced}
\end{figure}

When the condition $N>N_{cr}$ is fulfilled, the second phase corresponding to the stretched (open) 
chain emerges at relatively small extension of the spherical globule, 
$D-2R_0\sim N^0$, see  Eq.~(\ref{eq:tp1:D}). 
In the two-phase regime, the number of monomer units in the globular phase $n$ decreases approximately linearly with an increase in $D$, 
Eq.~(\ref{eq:ngl_tadpole_appr}). 
This leads to the growth of the positive term in $\mu_{globule}$, see Eq.~(\ref{eq:mu_globule_appr}) due to the surface effect, 
proportionally to $\gamma$. 
The reaction force $f$ weakly decreases with an increase in $D$, see Eq.~(\ref{eq:ftadpole_appr}), Figures \ref{fig:force_200}, and \ref{fig:force_500}. 
With an increase in $N$ or/and a deterioration of the solvent quality (i.e. increase in $\chi$), 
the range of stability of the two-phase state increases even in reduced coordinates $D/N$, Figure \ref{fig:force_reduced}, 
the negative slope of the ``quasi-plateau'' diminishes, and  the force-extension curve acquires the shape typical 
for systems undergoing a phase transition by passing via two-phase state (see Figure \ref{fig:FD_and_PV}~b). 

For the chains of finite length the quasi-plateau in the force-extension curves, Figures \ref{fig:force_200} and \ref{fig:force_500},
ends up with an abrupt drop of the reaction force. 
As it can be seen from analytical expressions, existence of the drop as well as the negative slope of the quasi-plateau 
is associated with the excess surface free energy of the globular phase. 
Positive contribution of the surface to the free energy of the two-phase (tadpole) state makes it at $D\geq D_2$ less advantageous 
than the one-phase state of the stretched open chain. 
According to Eq.~(\ref{eq:tp2_n_appr}), the minimal size of the globular head in the two-phase regime scales as 
$n\sim N^{3/4}$ and as it follows from the results of the quantitative theory, this value is quite large. 
This value of $n$ appreciably exceeds the number of monomers $N_{min}$ necessary to form a stable free globule. 
The latter value can be easily estimated from the condition that the free energy of the free globule, 
$F_{globule}=\mu N + \gamma R_0^2$, should be non-positive. This gives (with the account of  Eq.~(\ref{eq:R_0})):
\begin{equation}
	N \geq N_{min} = \frac{36\pi\gamma^3}{|\mu|^3 \varphi^2}.
\end{equation}

Remarkably in the long chain limit $(N \gg 1)$ the size of the minimal globule in the extended chain $n$ 
does not depend on the elastic constant $k$ of the open chain and its dependence on the solvent quality includes 
the same combination of partial parameters as in the case of the free globule:
\begin{equation}  
   n \sim \left(\frac{\gamma}{|\mu| \varphi^{2/3}}\right)^{3/4} N^{3/4} \quad , \quad
   N_{min} \sim \left(\frac{\gamma}{|\mu| \varphi^{2/3}}\right)^3.
\end{equation}
This combination of parameters also determines the quasi-plateau slope of the force-extension curve, Eq.~(\ref{eq:ftadpole_appr}).
As it can be see from Table 1, as $\chi$ increases, $n$ and $N_{min}$ decrease. 

Note that on $f(D)$ curves obtained in  the SCF modeling (Figures 5, 6) some peculiarity is observed, namely, a little peak in the vicinity of $D_1$ 
where the new phase corresponding to the stretched chain emerges. 
The monomer chemical potential in the extended phase, $\mu_{chain}$, 
is independent on the total number of monomer units in this phase, but depends on the force $f$, see Eq.~(\ref{eq:muchain_f})).
Therefore, the results of the quantitative theory lead to continuity of the force at this transition point.
An insufficient advantage of forming a tail (stretched chain phase) of very small size should lead to a certain 
widening of the stability range of the one-phase elongated globular state. 
It is also possible that the initial withdrawal of the stretched chain segment from the globule is accompanied by a 
change of the local shape of the globule surface.

More visible is the difference between the results of the theory and  the SCF modeling in the second transition point
which occurs at strong deformation, i.e. in the vicinity of $D_2$. 
According to the SCF results the transition from tadpole to uniformly stretched chain
is shifted towards stronger deformations, as compared to prediction of the theory.

This effect can be explained in the following way: At strong extensions (close to $D_2$), the number of monomer units $n$ 
in the globular head of the tadpole
is small and, as a result, the deformation affects not only its shape, but also the density and other partial characteristics,
similarly to that occcurs upon deformation of a small overcritical globule.  
As a consequence, force jump in the transition point observed in SCF caclulations 
turns out to be lower than that predcited by our theoretical model.

\subsection{Tadpole stability: about the role of the tail}
Let us recall that in the framework of the developed theory we study \emph{equilibrium} characteristics of the systems. 
Therefore we can equally consider both globule unfolding by increasing $D$ and globule refolding by decreasing $D$. 
In the latter case, the transition from the one-phase stretched chain state to the two-phase tadpole state 
occurs at extension $D_2$ and corresponds to formation of a new globular phase. 
Because of the interfacial tension the globular phase formation in the extended chain becomes possible 
only by drawing  a large number of monomer units $n > n_{min}$ into the globular phase. 
This jump-wise change in the number of monomer units in the stretched phase 
leads correspondingly to the jump in the reaction force at $D_2$ on the force-extension curves. 
\begin{figure}[t] 
  \begin{center}
    \includegraphics[width=12cm]{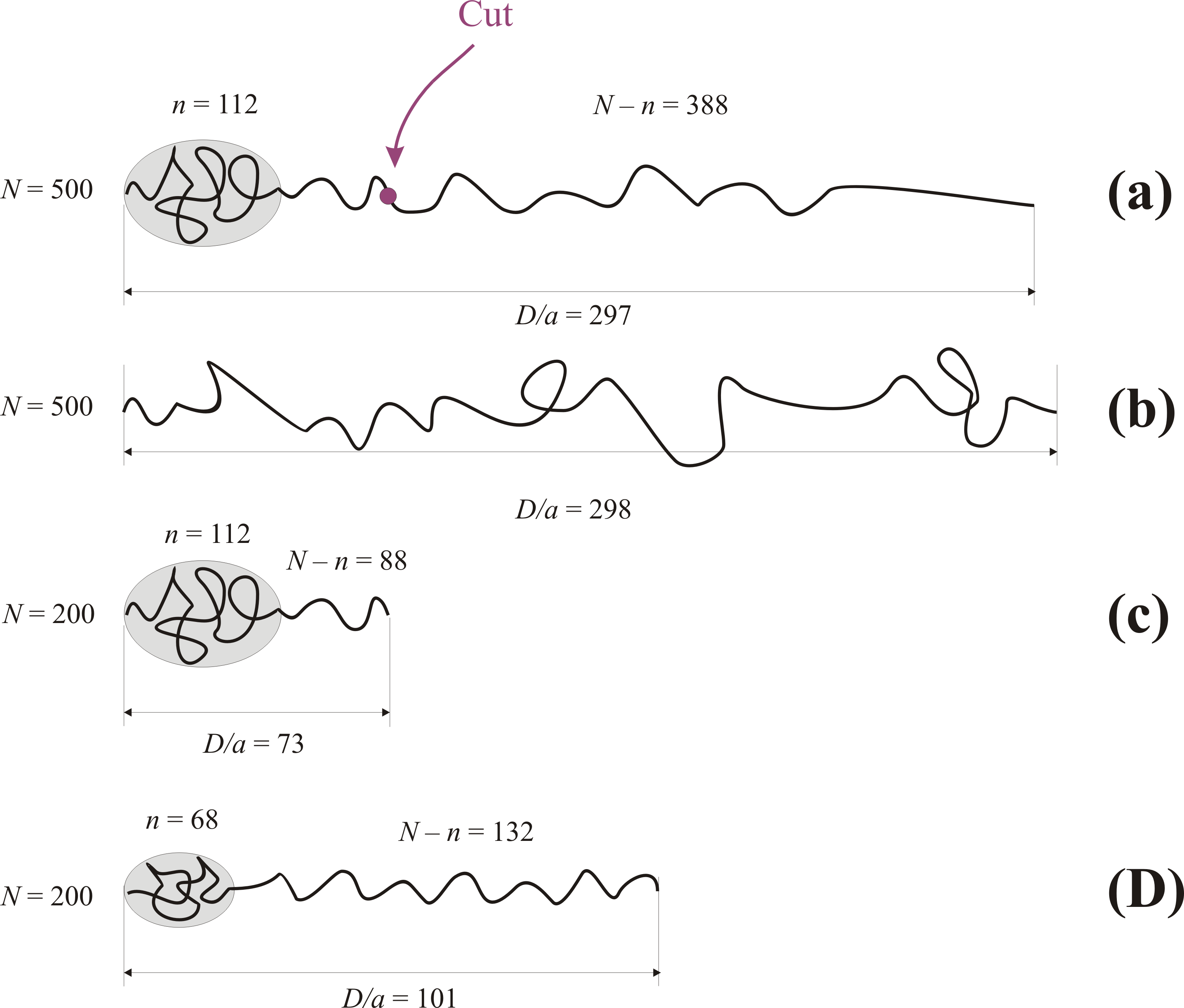}
  \end{center}
  \caption{Scheme of mental experiment.}
  \label{fig:mental}
\end{figure}

Analysis of the results of the theory and/or those of the SCF modeling shows another interesting effect. 
Consider the following mental experiment, Figure \ref{fig:mental}. 
Let us take a single polymer chain composed of $N=500$ monomer units and put it into a poor solvent with $\chi=2$ 
where it collapses into a dense ($\varphi a^3\approx 0.96$) spherical globule. 
Then we start an unfolding experiment: take the ends of the chain and increase the end-to-end distance $D$. 
According to the force-extension curve in Figure \ref{fig:force_500} and phase diagram in Figure \ref{fig:phd_D}, 
at $D/a \approx 11.3$ the mirophase segregated state appears, and the globule acquires the tadpole conformation. 
The tadpole can be extended up to $D/a \approx 297$, Figure \ref{fig:mental}~a,  
the globular head at this extension contains $n\approx 112$ monomers (more than 20 \% of the total amount of polymer) and 
its major axis is equal to $d/a \approx 7.3$. 
This is the limit of the tadpole stability: a larger extension leads to the tadpole breakup and to the transition into the stretched state, 
Figure \ref{fig:mental}~b. In other words, $n \approx 112$ is the smallest globular head that we can obtain by extending the globule 
for the given $N$ and $\chi$. 
Consider the 200th monomer of the chain: since the tail is extended uniformly, 
the position $D$ of this monomer can be found in a straightforward way: 
$D/a=7.3 + \frac{297-7.3}{500-112}\cdot(200-112)\approx 73$. Let us ``pin'' the 200th monomer at this position and cut (delete) 
the monomers from 201 to 500 inclusive, Figure \ref{fig:mental}~a, c. 
Now we have a globule containing $N=200$ monomers and extended at $D/a=73$ in the tadpole conformation, Figure \ref{fig:mental}~c. 
According to our ``cutting procedure'' the equilibrium conditions, Eq. (\ref{eq:eqconds}), 
will be fulfilled and no rearrangement in this ``reduced tadpole'' should occur. 
However, in contrast to the chain comprising 500 monomer units this tadpole can be further extended, from $D/a > 73$, up to $D/a \approx 101$.
The latter elongation corresponds to the limit of stability of the globule with $N=200$, where the head contains $n\approx 68$ monomers, Figure \ref{fig:mental}~d.


\section{Conclusions}
Motivated by recently obtained results of the SCF modeling~\cite{Polotsky:2009}, we have developed an analytical theory 
of mechanical unfolding of a homopolymer globule in the constant extension ensemble, i.e. by imposing the distance $D$ 
between the ends of the macromolecule. Correspondingly, the (conjugated) observable in this case is the average reaction force $f$.

Our approach consists in calculating and then comparing the free energies of the possible conformational states of the deformed globule. 
A weakly deformed globule is modeled by a prolate ellipsoid with constant density whereas at strong deformation 
it is represented as an ``open'' freely jointed chain. At moderate extensions, the co-existence of (weakly prolate) globular and (strongly) 
extended phases within a single macromolecule occur, i.e., the globule acquires a ``tadpole'' conformation with globular ``head'' and stretched ``tail''. 

Intramolecular microphase coexistence in the tadpole conformation implies equality of the monomer unit chemical potentials 
in two microphases and force balance between the globular head and the stretched tail which inevitably 
leads to asymmetry of the head's shape 

The model includes several parameters, such as degree of polymerization $N$, solvent strength quantified by Flory-Huggins parameter $\chi$, 
the mean polymer density in the unperturbed globule 
$\varphi$, interfacial tension coefficient $\gamma$, and monomer chemical potential in the globule $\mu$. 
In fact, setting $N$ and $\chi$ determines the values of $\varphi$, $\gamma$, and $\mu$. 
We have calculated the values of $\varphi$, $\gamma$, and $\mu$ numerically in the ``infinite globule approximation'' by using SF-SCF numerical approach. 
The values of $\varphi$ and $\mu$ can also be calculated using lattice Flory theory of polymer solutions; 
both approaches show very good agreement. The value of $\gamma$ can be found analytically for moderately poor solvent using Lifshitz theory for polymer globules~\cite{Lifshitz:1978, Ushakova:2006}.

Parameters of a random walk on the lattice for the freely jointed chain have been chosen similarly to those used in previous SCF 
numerical modeling~\cite{Polotsky:2009}. This allows us to compare force-extension curves obtained using SF-SCF modeling 
and in the framework of the developed theory. The results show not only a qualitative but also a very good quantitative agreement. 
This correspondence is better, the larger are $\chi$ and/or $N$.

Both the SCF calculations~\cite{Polotsky:2009} and the quantitative analytical theory prove that upon an increase in the end-to-end distance $D$, 
three regimes of deformation succesively occur: at small deformation the globule acquires a prolate shape, the reaction force grows 
linearly with the deformation, at moderate deformations the globule is in the tadpole conformation, 
reaction force is weakly decreasing, then at certain extension the globular head unravels and the globule completely unfolds, 
the reaction force drops down and then grows again upon further extension.

Furthermore, the analytical theory developed in the present paper makes it possible to go beyond the limits of the SCF calculations. 
First of all, it allows calculation of force-extension curves for large $N$, where the system size is large and numerical 
SCF calculations become very time and memory-consuming. Moreover, in the framework of the developed theory 
it is easy to calculate prolate globule-tadpole and tadpole-open chain transition points and to find corresponding conformational characteristics, 
such as the asymmetry and the number of monomers in the globular head, reaction forces and force jump at transition, in a wide range of $N$ and $\chi$. 
Assuming Gaussian elasticity for description of the stretched chain conformation we are able to obtain asymptotic 
analytical expressions 
for the force-extension curves in all the regimes of deformation as well as those for the globule characteristics in the transition points as a
function of the chain length $N$ and solvent strength $\chi$. 

Our analysis 
has shown that the system exhibits a \emph{critical point}, i.e. there exists a minimal chain length $N_{cr}$ below which, 
at $N < N_{cr}$, the intramolecular microphase segregation in the extended globule does not occur. 
The globule is deformed ``as a whole'', without intramolecular segregation but by progressive "dissolution" of its core . 
At the same time the value of $N_{cr}$ indicates the limit of applicability of our model since the assumption about constant polymer density in the globule
(``volume approximation'') is no more valid at smaller $N$, when the width of the globule's interfacial layer 
becomes comparable with the radius of the globular core.

Finally, we may conclude that the simple and physically transparent model developed in the present work 
adequately describes 
globule deformation in the constant extension ensemble.

\appendix

\section{Appendix. Determination of $\varphi$, $\mu$, and $\gamma$ by SCF approach}\label{app:parameters}
\begin{figure}[t] 
  \begin{center}
    \includegraphics[width=10cm]{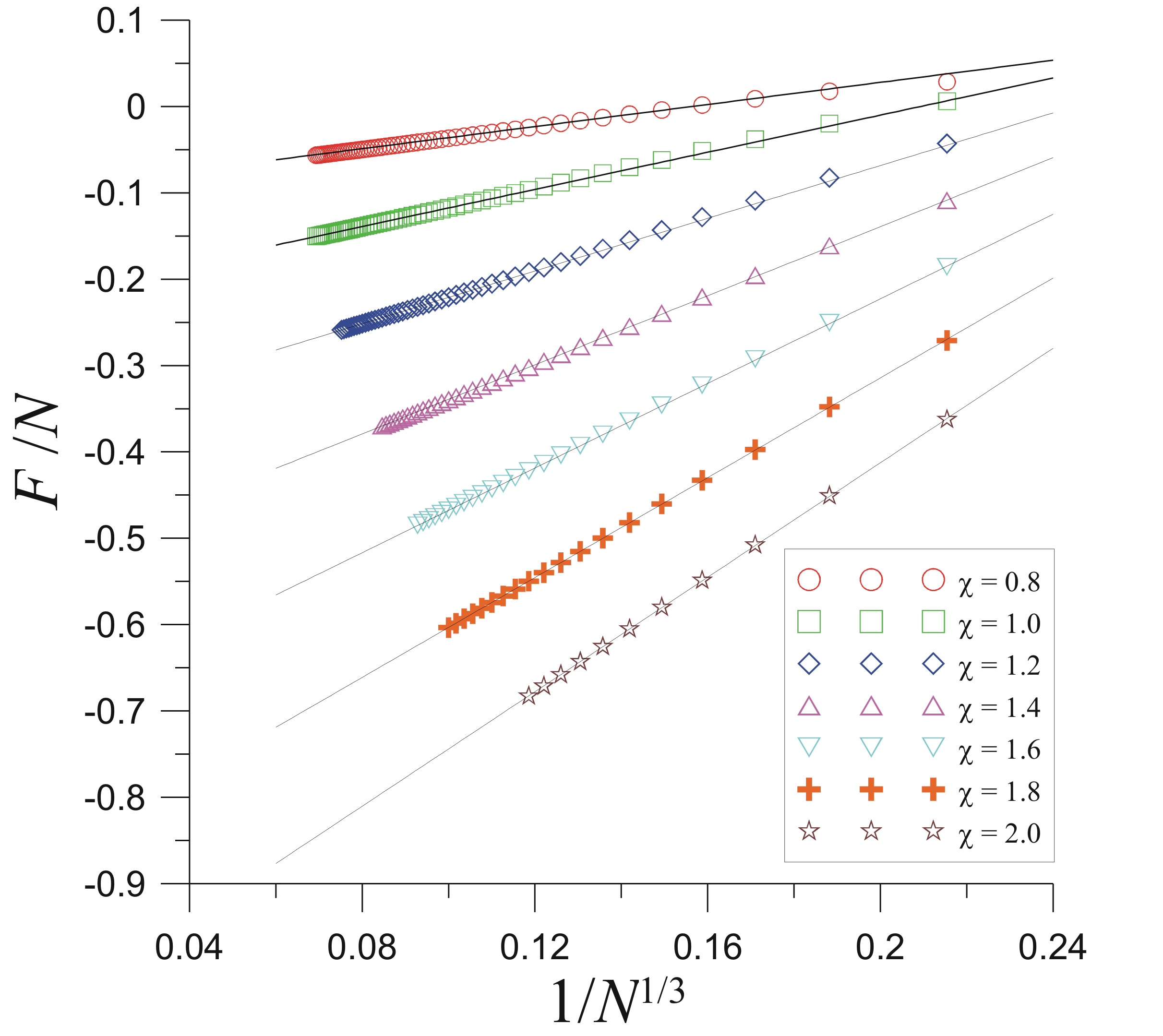}
  \end{center}
  \caption{Free energy of the unperturbed globule per one monomer as function of $1/N^{1/3}$ (symbols) and the corresponding
   linear fit (solid lines).}
  \label{fig:FE_globules}
\end{figure}

Consider a free (unperturbed) globule. It has a spherical shape and its free energy is given by
\begin{equation} \label{eq:Fsph1}
  F_{globule} = \mu N + \gamma S_{globule}
\end{equation}
If the polymer density within the globule is constant and equal to $\varphi$, then its radius and the surface area
\begin{equation}  \label{eq:Rglobule}
  R_{globule} = \left(\frac{3N}{4\pi\varphi}\right)^{1/3} \, , \quad 
  S_{globule} = 4\pi R^2_{globule} = 4\pi \left(\frac{3N}{4\pi\varphi}\right)^{2/3}
\end{equation}
and
\begin{equation} \label{eq:Fsph2}
  F_{globule} = \mu N + \gamma \cdot 4\pi \left(\frac{3N}{4\pi\varphi}\right)^{2/3}
\end{equation}
or, per monomer unit,
\begin{equation} \label{eq:Fsph3}
  \frac{F_{globule}}{N} = \mu + \gamma \cdot \left(\frac{36\pi}{\varphi^2}\right)^{1/3} N^{-1/3} .
\end{equation}
Hence, by performing SCF calculations on the 2-gradient cylindrical lattice~\cite{Polotsky:2009} for various $N$ and plotting the dependence $F_{globule}/N$ vs. $1/N^{1/3}$, Figure \ref{fig:FE_globules}, the values of $\mu$ (point of intersection with $F/N$ axis obtaining by extrapolation to $1/N^{1/3} = 0$) and $A = (36\pi/\varphi^2)^{1/3}\gamma$ can be found. To determine the interfacial tension coefficient $\gamma$, polymer density within the globule, $\varphi$ should be known.

It can be found with the aid of the axial distribution of monomer units  $n(z)$,
(the number of monomer units per $z$ axis unit):
\begin{equation} \label{eq:nz}
  n(z)=\sum_{r\geq a} L(r)\phi(r, z).
\end{equation}
where $L(r)$ is the number of lattice sites with coordinate $(r,z)$ and $\phi(r, z)$ is the polymer volume fraction profile 
obtained in 2-gradient SCF calculations.

In cylindrical coordinates, the surface of the spherical globule with the radius $R_{globule}$ is described by the equation 
$R_{globule}^2 = r^2 + z^2$
If the polymer density within the globule is constant, then $n(z)=\varphi\cdot \pi r^2 a = \varphi\cdot \pi (R_{globule}^2 - z^2)a = n_{max} - \varphi\cdot \pi z^2 a$ where 
\begin{equation}
	n_{max} = \varphi\cdot\pi R_{globule}^2 a = \varphi\cdot\pi \left(\frac{3N}{4\pi\varphi}\right)^{2/3} a
\end{equation}
is the peak value on the $n(z)$ profile. If $n_{max}$ is known, then $\varphi$ is easily found
\begin{equation}
	\varphi = \frac{16 n^3_{max}}{9\pi N^2 a^3}
\end{equation}

\begin{figure}[t] 
  \begin{center}
    \includegraphics[width=10cm]{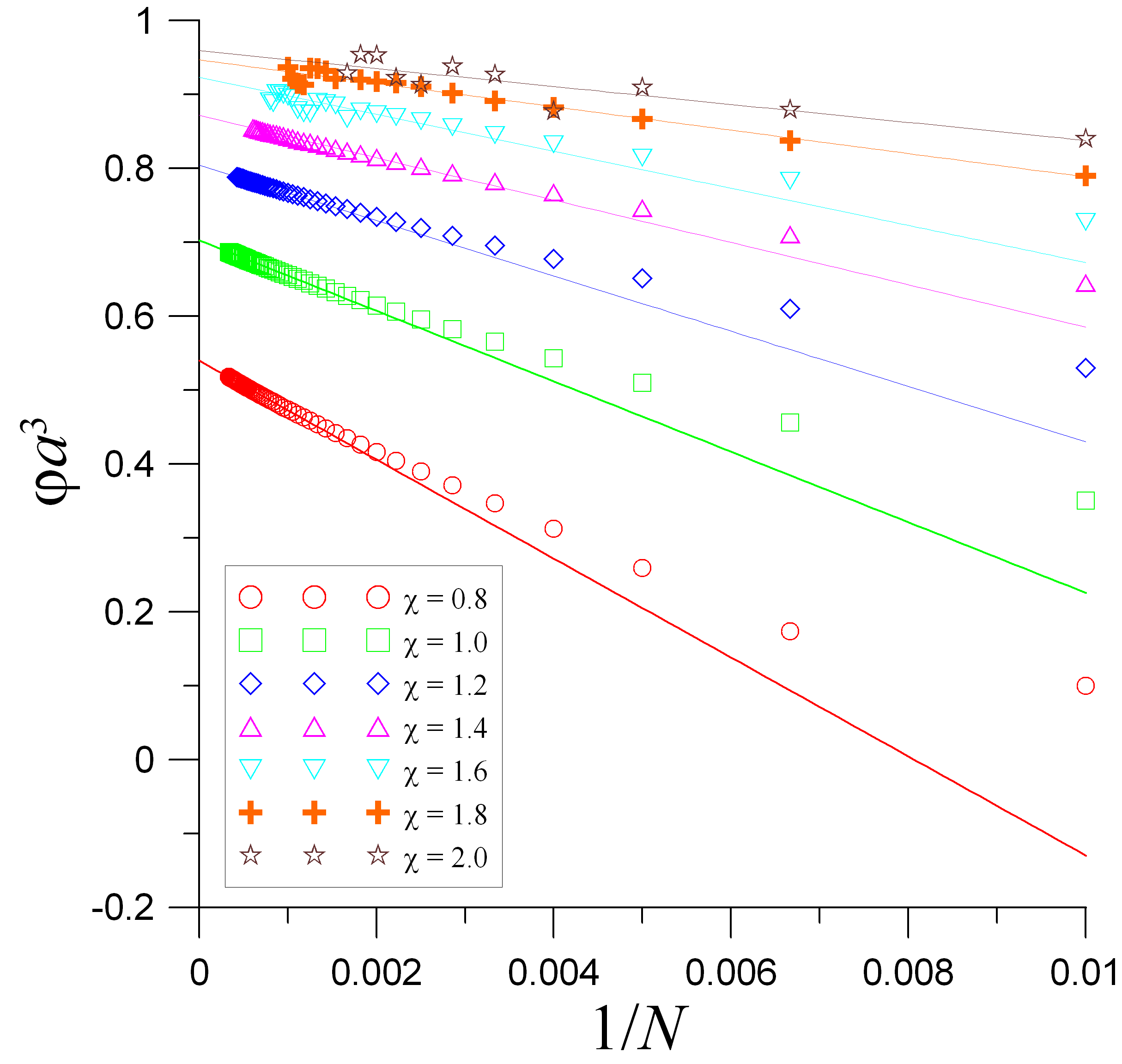}
  \end{center}
  \caption{Average polymer density in the unperturbed globule as function of $1/N$ and the corresponding extrapolation 
  to $N\to\infty$.}
  \label{fig:phi_globules}
\end{figure}

The value of $\varphi$ corresponding to the infinitely large globule can be found by extrapolating the dependence $\varphi(N)$ vs $1/N$ to $1/N \to 0$, Figure \ref{fig:phi_globules}. The values of $\mu$, $\gamma$, and $\varphi$ obtained using the above described procedure for the values of $\chi$ used in this work are shown  Table~\ref{table:mu_gamma_phi}. We find a very good agreement between the numerical results and those obtained in the framework of Flory theory, see Eqs~(\ref{eq:chi_tau}), (\ref{eq:gamma1_tau}) and Figure \ref{fig:mugammaphi}.

\begin{table}
\caption{Values of polymer concentration $\varphi$, monomer chemical potential $\mu$, and interfacial tension coefficient $\gamma$ calculated for different $\chi$ using SCF method}
 \label{table:mu_gamma_phi}
\centering
\begin{tabular}{|l|c|l|l|l|}
\hline
\hline
$\chi$ & $N$ & $\varphi a^3$ & $\mu$ & $\gamma a^2$  \\
\hline
			0.8 & 1000 $\div$ 3000  & 0.54 & -0.10 & 0.088\\
      \hline
			1.0 & 1000 $\div$ 3000  & 0.70 & -0.23 & 0.18\\
      \hline
			1.2 & 1000 $\div$ 2350  & 0.80 & -0.37 & 0.27\\
      \hline
			1.4 & 500 $\div$ 1650  & 0.87 & -0.54 & 0.38\\
      \hline
			1.6 & 500 $\div$ 1150  & 0.92 & -0.71 & 0.48\\
      \hline
			1.8 & 100 $\div$ 1000  & 0.94 & -0.89 & 0.58\\
      \hline
			2.0 & 100 $\div$ 600  & 0.96 & -1.08 & 0.670\\
      \hline
\hline
\end{tabular}
\end{table}

\section*{Acknowlegdement}
\label{Acknowlegdement}
Financial support by the Russian Foundation for Basic Research (RFBR)
through Project 08-03-00336a and by the Department of Chemistry and Marterial Science of the Russian Academy of Sciences is gratefully acknowledged.

\bibliographystyle{unsrt}
\bibliography{Globule}

\end{document}